\documentclass[a4paper]{article}
\usepackage{amssymb}

\usepackage{graphicx}
\usepackage{amsmath}
\usepackage{amssymb}
\usepackage{amsfonts}

\usepackage[a4paper]{geometry}

\input{tcilatex}

\begin{document}

\title{An effective 2D model for MHD\ flows with transverse magnetic field}
\date{February 9$^{th}$, 2000}
\author{A.Poth\'{e}rat$^{1}$, J.Sommeria$^{2}$, R Moreau$^{1}$ \\
$^{1}$Laboratoire EPM-MADYLAM (CNRS)\\
ENSHMG BP 95 38402 Saint Martin d'H\`{e}res Cedex.\\
$^{2}$Laboratoire de Physique (CNRS),\\
Ecole Normale Sup\'{e}rieure de Lyon,\\
46 all\'{e}e de l'Italie 69364 Lyon Cedex 07}
\maketitle

\begin{abstract}
This paper presents a model for quasi two-dimensional MHD flows between two
planes with small magnetic Reynolds number and constant transverse magnetic
field orthogonal to the planes. A method is presented that allows to take 3D
effects into account in a 2D equation of motion thanks to a model for the
transverse velocity profile. The latter is obtained by using a double
perturbation asymptotic development both in the core flow and in the
Hartmann layers arising along the planes. A new model is thus built that
describes inertial effects in these two regions. Two separate classes of
phenomena are thus pointed out : the one related to inertial effects in the
Hartmann layer gives a model for recirculating flows and the other
introduces the possibility of having a transverse dependence of the velocity
profile in the core flow. The ''recirculating'' velocity profile is then
introduced in the transversally averaged equation of motion in order to
provide an effective 2D equation of motion. Analytical solutions of this
model are obtained for two experimental configurations : isolated vortices
aroused by a point electrode and axisymmetric parallel layers occurring in
the MATUR (MAgneticTURbulence) experiment. The theory is found to give a
satisfactory agreement with the experiment so that it can be concluded that
recirculating flows are actually responsible for both vortices core
spreading and excessive dissipative behavior of the axisymmetric side wall
layers.
\end{abstract}

\section{Introduction.}

Magnetohydrodynamic flows at the laboratory scale have been the subject of
many investigations during the last decades, which lead to a rather good
level of understanding (see, for instance Hunt and Shercliff (1971) \nocite
{Hunt71} and Moreau (1990)\nocite{Moreau90}). In this paper, we focus on
flows of incompressible fluids, such as liquid metals, in the presence of a
uniform magnetic field $\mathbf{B}$. The magnetic Reynolds number $Rm=\mu
\sigma UL$ ($\mu $ denotes the fluid magnetic permeability, $\sigma $ its
electrical conductivity, $U$ and $L$ are typical velocity and length scales)
is supposed significantly smaller than unity, so that the actual magnetic
field within the fluid is close to $\mathbf{B}$. The fluid flows in a
container bounded by two insulating walls perpendicular to the magnetic
field (usually named \textit{Hartmann Walls }). Nothing is specified for the
other boundaries (for instance the wall parallel to the magnetic field) or
for the driving mechanisms (except when particular examples are considered).
The magnetic field is supposed high enough, so that both the Hartmann number
($Ha=aB\sqrt{\sigma /\rho \nu }$) and the interaction parameter ($N=\sigma
B^{2}a/\rho U$) are much larger than unity (here $a$ is the distance
separating the two Hartmann walls, $\rho $ the fluid density and $\nu $ its
kinematic viscosity). In such flows, the Hartmann boundary layers which
develop along the Hartmann walls are of primary importance.

One of the most important features of these flows is the fact that
turbulence is only weakly damped out by the electromagnetic force. Indeed,
because of their tendency to form quasi-two-dimensional (2D) structures,
these flows induce a significant current density only within the Hartmann
layers whose thickness is of the order $Ha^{-1}$. As a consequence the
quasi-2D core is only weakly affected by Joule dissipation and a highly
energetic turbulence may be observed (Lielausis (1975) \nocite{Lielausis75}%
). In such a configuration, the persistence of two-dimensional turbulence
and its specific properties have been found by Kolesnikov and Tsinober%
\textbf{\ }(1974)\nocite{Tsinober74} in decaying grid turbulence and then by
Sommeria (1986) in electromagnetically forced regimes.

To understand this persistence of turbulence and its
quasi-two-dimensionality the reader is referred to a number of earlier
papers. In particular, Alemany \textit{et al.}(1979) \nocite{alemany79}
demonstrated how an initially isotropic grid turbulence develops an
increasing anisotropy. However in this experiment, because there is no
confinement by Hartmann walls, the ohmic damping is of primary importance:
the characteristic time for both the development of the anisotropy and the
ohmic damping is $\rho /\sigma B^{2}$ and may be shorter than the eddy
turnover time. The key mechanisms are explained in Sommeria and Moreau
(1982) \nocite{Sommeria82} and in a review paper (Moreau 1998\nocite
{Moreau98}). More recently, Davidson (1997)\nocite{Dav97} pointed out the
crucial role of the invariance of the component of the angular momentum
parallel to the magnetic field (whereas the components perpendicular to $%
\mathbf{B}$ decrease on the timescale $\rho /\sigma B^{2}$) and Ziganov and
Thess (1998) \nocite{Ziganov98} achieved a numerical simulation of this
phenomenon exhibiting the sequences of events which lead to the formation of
column-like turbulent structures elongated in the direction of the magnetic
field. But these two theoretical approaches, as well as the experimental
part of Alemany \textit{et al.(1979)}, which do not involve the confinement
by Hartmann walls, are not directly relevant for the quasi-2D flows
considered here.

In this case, Sommeria and Moreau (1982) have described how the magnetic
field tends to suppress velocity differences in transverse planes. If the
Hartmann number and interaction parameter are sufficiently large, this
phenomenon can be considered as instantaneous so that the flow is not
dependent on the space coordinate associated with the field direction
anymore, except in Hartmann layers, where the velocity exhibits an
exponential profile given by the classical Hartmann layer theory.
Integrating the equation of motion along the field direction then provides a
2D Navier-Stokes equation with a forcing and a linear braking representing
electromagnetic effects and friction in the Hartmann layers. This ''2D core
model'' has provided a good quantitative prediction for various
electromagnetically driven flows (Sommeria, 1988). It has been generalized
by B\"{u}hler (1996)\nocite{Buhl96} to account for the presence of walls
with various conductivities, and applied to configurations of interest for
the design of lithium blankets in nuclear fusion reactors.

However, this 2D core model is only justified for $N$ and $Ha$ much larger
than unity, and discrepancies with experiments have been observed for
moderate values of the interaction parameter $N$. Then Ekman recirculating
flows are produced by inertial effects in the Hartmann layer. As a
consequence, a spreading of the vortex core was observed by Sommeria\ (1988) 
\nocite{Sommeria88} for vortices aroused by a point electrode. Such inertial
effects have been more systematically investigated in recent experiments of
electrically driven circular flows (Alboussi\`{e}re \textit{et al 1999} 
\nocite{Albouss99}). In the inertialess limit, complete 3D\ calculations
provide linear solutions for such flows or for parallel layers, but no
analytical model describes their non linear behavior due to inertial effects.

The present work aims at building such a model by a systematic expansion in
terms of the small parameters $Ha^{-1}$ and $N^{-1}$. The 2D core model of
Sommeria \& Moreau (1982) is recovered at the leading order, and
three-dimensional effects arise as perturbations.

In the next section we first recall the complete 3D equations. The
electromagnetic effects are interpreted as a diffusion of momentum along the
magnetic field direction, which tends to soften velocity differences between
transverse planes, thus driving the flow toward a 2D state in the core. We
also derive a 2D evolution equation for quantities averaged across the fluid
layer along the magnetic field direction (which we shall suppose
''vertical'' to simplify the description). This vertically averaged 2D
equation is always valid, even when the 2D core structure is not reached,
but it then involves terms depending on the vertical velocity profile,
similar to usual Reynolds stresses. For a 2D core with Hartmann boundary
layers, this vertically averaged equation reduces to the 2D core model of
Sommeria \& Moreau (1982), that we recall in section \textbf{2.2}. We stress
that it can be applied even in the parallel boundary layers near the lateral
walls, or in the core of a vortex electromagnetically driven around a point
electrode (scaling as $aHa^{-1/2}$ like parallel boundary layers). Indeed
the 2D core model compares well with linear theories involving a complete 3D
calculation.

Section \textbf{3} is devoted to the detailed investigation of the complete
3D equations, using a double perturbation method simultaneously in the core
and in the Hartmann layer. A first kind of 3D effects, discussed in section 
\textbf{3.2}, is the presence of recirculating flows driven by inertial
effects in the Hartmann layer. For axisymmetric flows, this is an Ekman
pumping mechanism. A second kind of 3D effect, occurring in the core, is
discussed in section \textbf{3.3 }: a perturbation of the 2D core, with a
profile quadratic in the vertical coordinate, is due to the finite time of
action of the electromagnetic diffusion of momentum along the vertical
direction. Thus in unsteady regimes, vortices are ''barrel'' shaped, instead
of truly columnar. Introducing some of these perturbations of the vertical
velocity profile in the vertically averaged equations yields an effective 2D
model, described in section \textbf{3.4}. This is the main result of the
present paper. The new terms involved in this model are mostly important for
small horizontal scales, leading in particular to new kinds of parallel
layers near curved walls or in the core of vortices, as specifically
discussed in section \textbf{3.5}.

This effective 2D model could be implemented in numerical computations of
various MHD flows between two Hartmann walls (or with a bottom wall and a
quasi-horizontal free surface). We discuss in section \textbf{4} the
application to axisymmetric flows. We apply the results to the
electromagnetically generated vortex of Sommeria (1988) and to the MATUR
experiments (Alboussi\`{e}re \textit{et al. 1999}). The discrepancies of the
2D core model are reasonably accounted by our effective 2D model, taking
into account the influence of recirculating flows.

\section{General equations and 2D-core model.}

\subsection{General equations and z-averaging}
\begin{figure}
\centering
\includegraphics[width=0.7\textwidth]{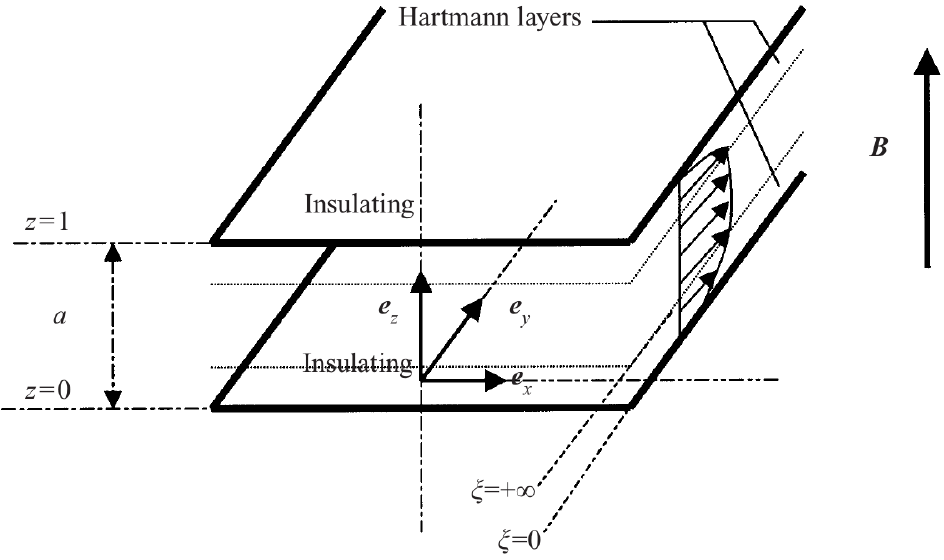}
\caption{\label{Geometric configuration} Geometric configuration
considered in our model.}
\end{figure}
The fluid of density $\rho $, kinematic viscosity $\nu $ and electrical
conductivity $\sigma $ is supposed to flow between two electrically
insulating plates orthogonal to the uniform magnetic field $\mathbf{B}$ (see
figure \ref{Geometric configuration}). We suppose $\mathbf{B}$ is vertical
for the simplicity of description (although there is no gravity effect). We
start from the Navier-Stokes equations for an incompressible fluid with 
\textit{a priori} 3D velocity field $\mathbf{u}$ and pressure $p$. The
non-dimensional variables and coordinates are defined from physical
variables (labelled by the subscript ()$_{dim}$as 
\begin{equation}
\begin{array}{cccc}
x_{dim}=\dfrac{a}{\lambda }x & t_{dim}=\dfrac{a}{\lambda U}t & \mathbf{j}%
_{\bot dim}\mathbf{=}\sigma BU\mathbf{j} & p_{dim}=\rho U^{2}p \\ 
y_{dim}=\dfrac{a}{\lambda }y & \mathbf{u_{\perp }}_{dim}\mathbf{=}U\mathbf{%
u_{\perp }} & j_{zdim}\mathbf{=}\lambda \sigma BUj_{z} & \mathbf{B}_{dim}%
\mathbf{=}B\mathbf{e}_{z} \\ 
z_{dim}=az & w_{dim}=\lambda Uw &  & 
\end{array}
\label{Characteristic values}
\end{equation}
Note that we distinguish the scales parallel and perpendicular (with the
aspect ratio $\lambda $) to the magnetic field, and the corresponding
velocities ($\mathbf{u_{\perp },}w$\textbf{)} and currents ($\mathbf{j}%
_{\bot },j_{z}$) accordingly. The subscript $\bot $ denotes the vector
projection in the direction perpendicular to the magnetic field. The
Hartmann number $Ha$ and the interaction parameter $N$ are defined as :
\begin{equation}
\begin{array}{cc}
Ha=aB\sqrt{\dfrac{\sigma }{\rho \nu }}, & N=\dfrac{\sigma B^{2}a}{\rho U}
\end{array}
.  \label{Main non dim numbers}
\end{equation}
Notice that the Reynolds number is defined as $Re=Ha^{2}/N$. It may be
noticed that all these non-dimensional numbers are built with the layer
thickness $a$.

Using these dimensionless variables, the motion equations write

\begin{eqnarray}
\mathbf{\nabla }_{\perp }.\mathbf{u_{\perp }+}\partial _{z}w&=&0\text{,}
\label{adim continuity in the core} \\
\dfrac{\lambda }{N}\left( \partial _{t}\mathbf{u_{\perp }}+\mathbf{u_{\perp
}.\nabla }_{\perp }\mathbf{u_{\perp }}+w\partial _{z}\mathbf{u_{\perp }}%
+\nabla _{\perp }p\right) -\dfrac{\lambda ^{2}}{Ha^{2}}\Delta _{\perp }%
\mathbf{u_{\perp }-}\dfrac{1}{Ha^{2}}\partial _{zz}^{2}\mathbf{u_{\perp }}&=&%
\mathbf{j}_{\bot }\times \mathbf{e}_{z},  \label{adim NS in the core} \\
\dfrac{\lambda }{N}\left( \partial _{t}w+\mathbf{u_{\perp }.\nabla }_{\perp
}w+w\partial _{z}w+\partial _{z}p\right) -\dfrac{\lambda ^{2}}{Ha^{2}}\Delta
_{\perp }w\mathbf{-}\dfrac{1}{Ha^{2}}\partial _{zz}^{2}w&=&0,
\label{adim NS  vertical in the core}
\end{eqnarray}
\begin{equation}
\mathbf{\nabla }_{\perp }.\mathbf{j}_{\bot }+\partial _{z}j_{z}=0,
\label{current continuity core}
\end{equation}

\begin{equation}
\mathbf{j}=-\mathbf{\nabla }\phi +\mathbf{u}\times \mathbf{e}_{z}.
\label{Ohm'slaw}
\end{equation}
The electromagnetic force $\mathbf{j}\times \mathbf{e}_{z}$ has been
included, where the electric current density\textbf{\ }$\mathbf{j}$ is
related to the\ electric potential $\phi $ by (\ref{Ohm'slaw}), representing
Ohm's law. As the action of the induced magnetic field is negligible, the
electromagnetic equations reduce to the condition of divergence-free current
(\ref{current continuity core}).

The electromagnetic force depends linearly on the velocity field, but in a
non-local way. The current density $\mathbf{j}$ can be eliminated in (\ref
{adim NS in the core}) (see for example Roberts (1967) \nocite{Roberts67}).
Denoting $\mathbf{j}\times \mathbf{e}_{z}\mathbf{=f+\nabla }p_{\varepsilon }$
in order to distinguish the rotational part and the divergent part of the
Lorentz force, taking twice the curl of $\mathbf{j}\times \mathbf{e}_{z}$
and using (\ref{current continuity core}) and (\ref{Ohm'slaw}) yields :

\begin{equation}
\Delta \mathbf{f}=\partial _{zz}^{2}\mathbf{u,}
\label{Electromagnetic action equation}
\end{equation}
Note that $p_{\varepsilon }$ can be included in the pressure term. In the
limit of strong magnetic field, the force becomes very large, resulting in a
fast damping by Joule effect, except if $\partial _{zz}^{2}\mathbf{u}$ is
small, \textit{i.e.} the flow is close to two-dimensional. In this case, $%
\Delta \mathbf{f\simeq }\Delta _{_{\bot }}\mathbf{f}$ , where $\Delta
_{_{\bot }}$ stands for the Laplacian in the plane perpendicular to the
magnetic field. Sommeria \& Moreau (1982) proposed to interpret this force
as a momentum diffusion along the direction of the magnetic field, with a
''diffusivity'' $\tfrac{\sigma B^{2}a^{2}}{\lambda ^{2}\rho }$ depending on
the transverse scale $a/\lambda $. This diffusion tends to achieve
two-dimensionality in the fluid interior when the corresponding diffusion
time is smaller than the eddy turnover time $\tfrac{a}{\lambda U},$ \textit{%
i.e.}:

\begin{equation}
\dfrac{\rho }{\sigma B^{2}}\lambda ^{2}<<\dfrac{a}{\lambda U}\text{\textit{%
i.e }}\lambda ^{3}<<N,  \label{difftime}
\end{equation}

However in order to take into account weak 3D effects, we shall not assume
two-dimensionality right away, but get a 2D model by integrating the 3D
equations along the direction of the magnetic field (\textit{i.e.} the $z$
coordinate), leading to a 2D dynamics for z-averaged quantities. We define
the z-average of any quantity $g$ and its departure from average $g\prime $
respectively by

\begin{equation}
\begin{array}{cc}
\bar{g}(x,y)=\int_{0}^{1}gdz, & g^{\prime }(x,y,z)=g-\bar{g}
\end{array}
.  \label{average-departure}
\end{equation}
The z--average of the momentum equation (\ref{adim NS in the core}) then
leads to

\begin{equation}
\dfrac{\lambda }{N}\left( \partial _{t}\mathbf{\bar{u}}_{\bot }+\left( 
\mathbf{\bar{u}}_{\bot }\mathbf{.\nabla }\right) \mathbf{\bar{u}}_{\bot }%
\mathbf{+}\overline{\left( \mathbf{u}_{\bot }^{\prime }\mathbf{.\nabla }%
\right) \mathbf{u}_{\bot }^{\prime }}+\mathbf{\nabla }\overline{p}\right) =%
\dfrac{\lambda ^{2}}{Ha^{2}}\mathbf{\Delta \bar{u}}_{\bot }+\dfrac{1}{Ha^{2}}%
\mathbf{\tau }_{W}+\overline{\mathbf{j}}\mathbf{\times e}_{z}
\label{2d Integrate electromagnetic motion equation}
\end{equation}
for each velocity component $\mathbf{u}_{\bot }$ ($j\in \left\{ 1,2\right\} $%
) perpendicular to the magnetic field. Here $\mathbf{\tau }_{W}=-[\partial
_{z}\mathbf{u}_{\bot }$ $(z=0)-\partial _{z}\mathbf{u}_{\bot }(z=a)]$
denotes the\ sum of the non-dimensional viscous stresses at the lower and
upper walls. The z-average of the continuity equation (\ref{adim continuity
in the core}), with the impermeability conditions at the walls, indicates
that the z-averaged velocity is divergence free in two dimensions. Therefore
the initial 3D problem translates into a problem of an incompressible flow $%
\mathbf{\bar{u}}_{\bot }$ satisfying the 2D Navier-Stokes equation with two
added terms : the divergence of a Reynolds stress tensor $\underline{%
\underline{\mathbf{\nabla }}}.\overline{\mathbf{u}_{\bot }^{\prime t}\mathbf{%
u}_{\bot }^{\prime }}$, resulting from the momentum transport by the 3D flow
component, and the wall friction term $\mathbf{\tau }_{W}.$ The knowledge of
both terms requires a model for the vertical velocity profile whose
derivation is the main issue of section \textbf{3}.

The electromagnetic term $\overline{\mathbf{j}}\mathbf{\times e}_{z}$ can be
expressed from the current density $j_{W}$ $(x,y)$ injected in the fluid
through the two walls (at $z=0$ and $z=1$).\ Indeed, the z-average of (\ref
{current continuity core}) yields $\mathbf{\nabla }_{\bot }.\overline{%
\mathbf{j}_{\bot }}=j_{W}$, and the z-average of (\ref{Ohm'slaw}) yields $%
\mathbf{\nabla }_{\bot }\times \overline{\mathbf{j}_{\bot }}=0$ \ (using the
incompressibility condition $\mathbf{\nabla }_{\bot }.\mathbf{\bar{u}}_{\bot
}=0$). Thus the z-averaged current can be expressed as the gradient of a
scalar $\Psi _{0}$ satisfying a Poisson equation,

\begin{equation}
\begin{array}{cc}
\overline{\mathbf{j}_{\bot }}=\dfrac{1}{Ha}\mathbf{\nabla }\Psi _{0}, & 
\dfrac{1}{Ha}\Delta _{\bot }\Psi _{0}=-j_{W}
\end{array}
.  \label{current}
\end{equation}
We shall consider either the case of insulating walls or the case of a
current density imposed on electrodes (in more complex cases of conducting
Hartmann walls, $j_{W}$ would be determined by a matching with Ohm's law in
the conductor). The boundary conditions on the side walls for $\Psi _{0}$
depend on the electrical condition : for electrically insulating lateral
walls (supposed tangent to the magnetic field), there is no normal current,
so that the normal derivative of $\Psi _{0}$ vanishes (Neuman conditions).
By contrast, for a perfectly conducting lateral wall , the current is
normal, so that $\Psi _{0}$ is constant on the wall (Dirichlet conditions)

Using (\ref{current}), the electromagnetic force $\overline{\mathbf{j}}%
\mathbf{\times e}_{z}$ in (\ref{2d Integrate electromagnetic motion equation}%
) can be expressed as a divergence-free horizontal vector, and the 2D
equation of motion writes :

\begin{equation}
\dfrac{\lambda }{N}\left( \partial _{t}\mathbf{\bar{u}}_{\bot }+\left( 
\mathbf{\bar{u}}_{\bot }\mathbf{.\nabla }\right) \mathbf{\bar{u}}_{\bot }%
\mathbf{+}\overline{\left( \mathbf{u}^{\prime }\mathbf{.\nabla }\right) 
\mathbf{u}^{\prime }}+\mathbf{\nabla }\overline{p}\right) =\dfrac{\lambda
^{2}}{Ha^{2}}\mathbf{\Delta }_{\bot }\mathbf{\bar{u}}_{\bot }+\dfrac{1}{%
Ha^{2}}\mathbf{\tau }_{W}+\dfrac{1}{Ha}\mathbf{u}_{0},
\label{2D integrate motion equation}
\end{equation}
where the 2D velocity field$\ \mathbf{u}_{0}\mathbf{\ }$is defined as $%
\mathbf{u}_{0}=\mathbf{\nabla }\Psi _{0}\times \mathbf{e}_{z}$.

\subsection{The 2D core model.}

\subsubsection{The Hartmann friction}

In the boundary layers, the z-derivatives dominate in (\ref{adim continuity
in the core}-\ref{Ohm'slaw}), resulting in the Hartmann velocity profile,
near the wall $z=0$

\begin{equation}
\mathbf{u}_{\perp }=\mathbf{u}^{-}\left( 1-e^{-Ha\text{ }z}\right) ,
\label{0 order Hartmann layer velocity profile}
\end{equation}
where $\mathbf{u}^{-}$ is the horizontal velocity near the wall, but outside
the boundary layer. The corresponding wall stress is :

\begin{equation}
\mathbf{\tau }^{-}=-Ha\text{ }\mathbf{u}^{-}.  \label{Hartmann theory stress}
\end{equation}
At the wall $z=a$ we shall consider either a free surface, supposed
horizontal, with no stress, either a solid wall, with corresponding velocity 
$\mathbf{u}^{+}$ and wall stress $\mathbf{\tau }^{+}.$

We consider for the moment a 2D core velocity, so that $\mathbf{u}^{\mathbf{+%
}}=\mathbf{u}^{\mathbf{-}}\simeq \overline{\mathbf{u}}$ (neglecting the
velocity fall in the boundary layer, as the latter is thin ($a/Ha)$ compared
with the total thickness $a$). This wall stress introduces a global linear
braking with characteristic time (for one Hartmann layer)

\begin{equation}
t_{H}=\dfrac{a^{2}}{\nu }\dfrac{1}{Ha}  \label{damping time expression}
\end{equation}
and the 2D core velocity field satisfies in non-dimensional form :

\begin{equation}
\dfrac{\lambda }{N}\left[ (\partial _{t}+\mathbf{\bar{u}}_{\bot }.\mathbf{%
\nabla })\mathbf{\bar{u}}_{\bot }+\mathbf{\nabla }\bar{p}\right] =\dfrac{%
\lambda ^{2}}{Ha^{2}}\mathbf{\Delta \bar{u}}_{\bot }+\dfrac{1}{Ha}\left( 
\mathbf{u}_{0}-n\mathbf{\bar{u}}_{\bot }\right) ,
\label{0-ordrer integrate equation}
\end{equation}
where $n$ is the number of Hartmann walls ($n=1$ in the case with a free
surface and $n=2$ for a flow between two Hartmann walls, such that the
friction is doubled).

The whole model was discussed by Sommeria and Moreau (1982) and applied to
various cases. It applies for sufficiently large perpendicular scales $%
\dfrac{a}{\lambda }$, such that condition (\ref{difftime}) is satisfied. In
principle it should break down in the parallel boundary layers, of thickness 
$\mathcal{O}(aHa^{-1/2})$, but it is interesting to test its validity in
this case. We shall consider two cases for which a three-dimensional
analytical solution is available as a reference: the parallel side boundary
layer and an isolated vortex aroused by a point-electrode.

\subsubsection{Sidewall layers}
\begin{figure}
\centering
\includegraphics[width=0.7\textwidth]{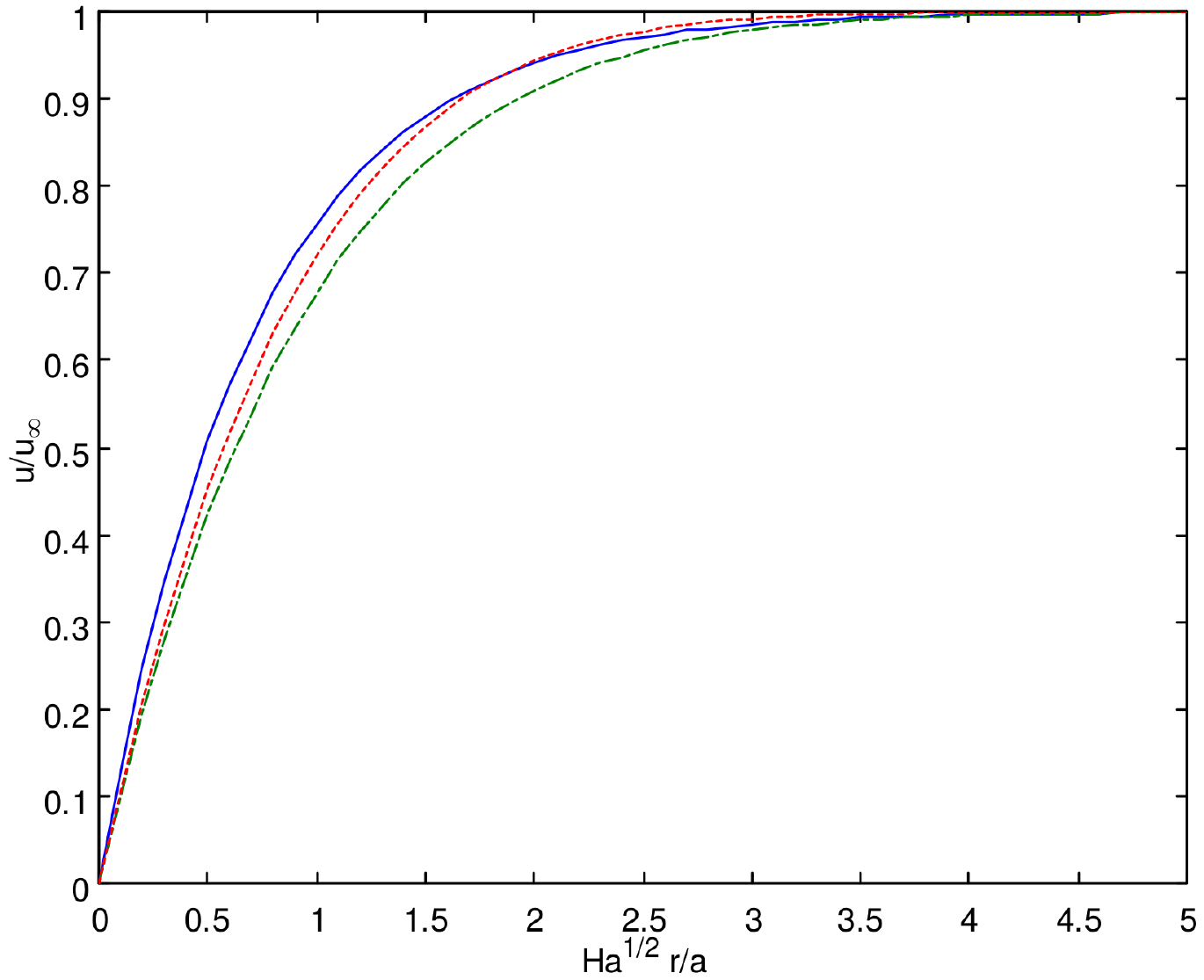}
\caption{\label{2D/3D paralell layer profiles} Comparison between the
1D profile (16) (solid line) and the corresponding profile for the 2D
averaged solution of Shercliff (1953) : the dotted line represents the
profile at $z=1/2$ and the dashed line, the z average profile.}
\end{figure}
Let us consider the case of a duct flow with rectangular section, as first
solved by Shercliff (1953)\nocite{Shercliff53}. In this case, the flow is
driven by pressure drop, which can be modelled with a uniform forcing
velocity $\mathbf{u}_{0}$ (like in the case of a uniform electromagnetic
driving by a transverse current). Then, equation (\ref{0-ordrer integrate
equation}) reduces to

\begin{equation}
\dfrac{\lambda ^{2}}{Ha}\partial _{yy}\bar{u}_{x}\left( y\right) -\left( 2%
\bar{u}_{x}\left( y\right) -u_{0x}\right) =0,  \label{2D paralellel layer}
\end{equation}
the solution of which is, near the side wall supposed located at $y=0$ :

\begin{equation}
\dfrac{\bar{u}\left( y\right) }{u_{0x}}=1-\exp \left( -y\dfrac{\sqrt{2Ha}}{%
\lambda }\right) .  \label{2D para layer velocity profile}
\end{equation}

Notice that this side boundary layer has a thickness $\tfrac{a}{\lambda }%
\simeq \tfrac{a}{\sqrt{Ha}}$ which results from a balance between lateral
diffusion, with time scale $\tfrac{a^{2}}{\lambda ^{2}\nu }$, and the
Hartmann friction with time scale $t_{H}$ ($\tfrac{a}{\lambda }\simeq \sqrt{%
\nu t_{H}}$). The velocity profile is plotted in figure (\ref{2D/3D paralell
layer profiles}) and compared with the 3D solution (see for instance Moreau%
\textit{\ }(1990)). \nocite{Moreau90}\textit{\ }\ Both the velocity in the
middle plane ($z=1/2$) and the z-averaged velocity are found in reasonable
agreement with the 2D core model although the hypotheses the latter relies
on are not fully satisfied in these side boundary layers. The profiles of
figure (\ref{Paralelle layer velocity profiles at z=cte.}) confirm that the
3D solution is not very far for from a 2D core. Notice that the electric
condition at the parallel wall is not of great importance since it only
induces a variation of a few percents on the velocity. By contrast, the
Hartman wall has to be insulating as discussed in section \textbf{3.1 }:
indeed, with conducting walls , there would be strong jets in the parallel
layer which cannot be described by this model.

\subsubsection{Isolated vortices}
\begin{figure}
\centering
\includegraphics[width=0.7\textwidth]{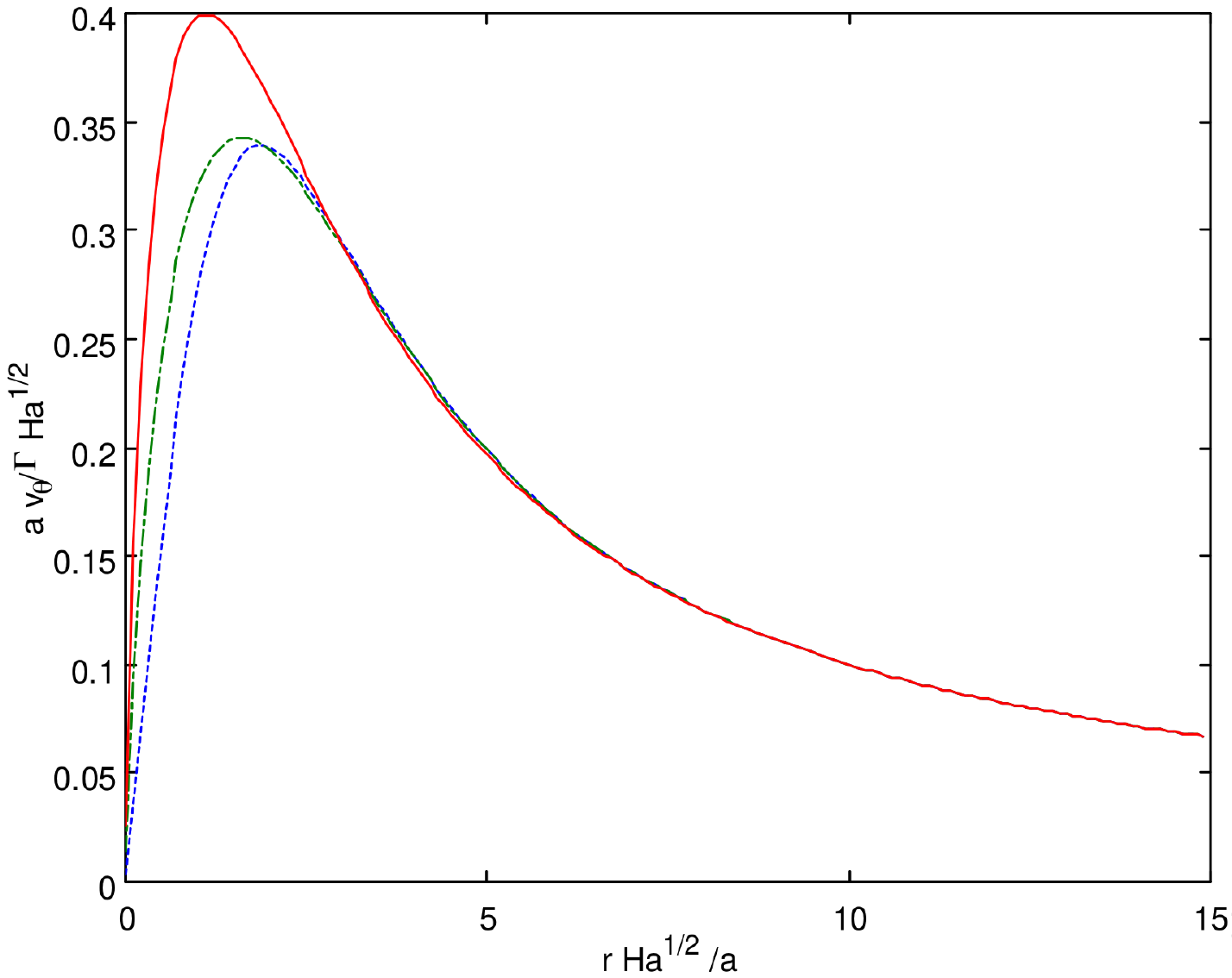}
\caption{\label{2D/3D linear vortex profile} Comparison between the
1D velocity profile (18) (solid line) and the 2D\ solution of Sommeria
(1988) for an electrically driven vortex. The dotted line represents the
profile at $z=1/2$ and the dashed line, the average profile.}
\end{figure}
Here, the 2D model is used to compute the velocity profile for an isolated
vortex driven by the electric current injected at a point electrode located
in the bottom plate, experimentally studied by Sommeria (1988). The upper
surface is free (but remaining quasi-horizontal)\ and side walls are
supposed very far. Therefore the source term $j_{W}$ in (\ref{current})%
\textbf{\ }is a Dirac\textbf{\ }function with integral equal to the injected
current $I$, and the corresponding forcing \textbf{\ }$\mathbf{u}_{0}$%
\textbf{\ }is azimuthal and depends on the radius $r$ as

\begin{equation}
u_{0\theta }=\dfrac{\Gamma }{r}\text{\ avec }\Gamma =\dfrac{I}{2\pi \sqrt{%
\rho \sigma \nu }},  \label{pointnelectrode circulation}
\end{equation}
where the velocity and the space coordinate have been rescaled using $U=%
\tfrac{\Gamma \sqrt{Ha}}{a}$ and $\lambda =\sqrt{Ha}$ (which corresponds to
the non-dimensional parallel layer thickness). The radial velocity profile
then results from the balance between electric forcing , Hartmann braking,
and lateral viscous stress. A steady laminar and axisymmetric solution of (%
\ref{0-ordrer integrate equation}) in polar coordinates is given by

\begin{equation}
u_{\theta }=\dfrac{1}{\tilde{r}}-K_{1}\left( \tilde{r}\right) ,
\label{2D Point elec velocity profile}
\end{equation}
where $K_{1}$ denotes the modified Bessel function of the second kind. Hunt
and\ Williams\textit{\ }(1968)\nocite{Hunt68} have performed a complete
asymptotic three-dimensional resolution of an analogous problem for large
values of the Hartmann number. The latter can be adapted to the present case
through simple transformations (Sommeria\textit{\ }1988) :

\begin{equation}
v_{\theta }=\dfrac{1}{r}\left( 1-\dfrac{1}{2}\exp \left( \dfrac{-r^{2}}{4z}%
\right) -\dfrac{1}{2}\exp \left( \dfrac{-r^{2}}{4\left( 2-z\right) }\right)
\right) .  \label{3D vortex profile}
\end{equation}
As in the previous sub-section,\ we compare the value of this solution at
the middle plane ($z=1/2$) and its z-average to the 2D solution (\ref{2D
Point elec velocity profile})(see figure \ref{2D/3D linear vortex profile}).
A reasonable agreement between 2D and 3D theories is obtained, in spite of
the very singular behavior of the 3D solution near the electrode. It is
interesting to notice that the simplified 2D theory gives the right orders
of magnitude for the core diameter and the maximum velocity as well.

\section{An effective quasi 2-D model.}

\subsection{Non-dimensional basic equations.}

In this section the modifications of the Hartmann profile (\ref{0 order
Hartmann layer velocity profile}) are derived by a perturbation method. Let
us first replace equation (\ref{Ohm'slaw})\textbf{\ }by its curl $\mathbf{%
\nabla \times j}=\left( \mathbf{e}_{z}.\mathbf{\nabla }\right) \mathbf{u}$
in order to express the electromagnetic effects directly in terms of the
velocity field. Distinguishing the transverse and parallel components of $%
\mathbf{\nabla \times j,}$ we get the equations for $\mathbf{j}$ :

\begin{eqnarray}
\mathbf{\nabla }_{\bot }\mathbf{\times j}_{\bot }&=&\partial _{z}w,
\label{curl of j} \\
\mathbf{e}_{z}\times \partial _{z}\mathbf{j}_{\bot }-\lambda ^{2}\left( 
\mathbf{e}_{z}\times \mathbf{\nabla }_{\perp }\right) j_{z}&=&\partial _{z}%
\mathbf{u_{\perp }.}  \label{adim Rot of the Ohm's law in the core}
\end{eqnarray}

We could deduce from these three equations the equation for the force $%
\mathbf{j}_{\bot }\times \mathbf{e}_{z}$ (the non dimensional form of (\ref
{Electromagnetic action equation})), but the information on the boundary
condition for $j_{z}$ would be lost.

In the Hartmann boundary layer, the $z$ coordinate scales like the Hartmann
layer thickness $a/Ha$. We use the subscript $(\ )_{h}$ to denote the
variables within the boundary layer which are functions of the argument$\
\xi $ $=Ha$ $z$ , the stretched $z$ coordinate ($\mathbf{u}_{h}$ and $w_{h}$
denote the velocity components perpendicular and parallel to the magnetic
field, respectively, whereas $\mathbf{j}_{h}$ and $j_{\xi }$ stand for the
horizontal and vertical electric current density). Then, equations (\ref
{curl of j}) and (\ref{adim Rot of the Ohm's law in the core}) become : 
\begin{eqnarray}
\dfrac{1}{Ha}\mathbf{\nabla }_{\perp }\times \mathbf{j}_{h}=\partial _{\xi
}w_{h},  \label{curl of j in the Hartmann layer} \\
-\dfrac{\lambda ^{2}}{Ha}\left( \mathbf{e}_{z}\times \mathbf{\nabla }_{\bot
}\right) j_{\xi }=-\mathbf{e}_{z}\times \partial _{\xi }\mathbf{j}%
_{h}+\partial _{\xi }\mathbf{u}_{h}.
\label{divergence of laplace force in the Hartmann layer}
\end{eqnarray}
They have to be completed by the condition of conservation of electric
current (\ref{current continuity core}) which becomes within the Hartmann
layer : 
\begin{equation}
\dfrac{1}{Ha}\mathbf{\nabla }_{\perp }.\mathbf{j}_{h}=-\partial _{\xi
}j_{\xi }.  \label{non dim current continuity in the Hartmann layer}
\end{equation}
With the same transformation, the equation of motion (\ref{adim continuity
in the core}-\ref{adim NS in the core}) become\footnote{%
The $z$ component of the momentum equation, omitted here, just states that
the pressure is independent of $z$ to a precision at least of the order $%
1/Ha $.} :

\begin{equation}
\xi =Ha\text{ }z  \label{H-layer Characteristic High}
\end{equation}

\begin{eqnarray}
\dfrac{1}{Ha}\mathbf{\nabla }_{\perp }.\mathbf{u}_{h}&=&\partial
_{\xi }w_{h},  \label{non dim continuity equation in the hartmann layer} \\
\dfrac{\lambda }{N}\left( \partial _{t}\mathbf{u}_{h}+\mathbf{u_{h}.\nabla }%
_{\perp }\mathbf{u}_{h}+Haw_{h}\partial _{\xi }\mathbf{u}_{h}+\nabla _{\perp
}p_{h}\right) -\dfrac{\lambda ^{2}}{Ha^{2}}\Delta _{\perp }\mathbf{u}%
_{h}&=&\partial _{\xi \xi }^{2}\mathbf{u}_{h}\mathbf{+j}_{h}\times \mathbf{e}%
_{z}.  \label{non dim NS in the hartmann layer}
\end{eqnarray}
%
The boundary conditions to be satisfied by the solutions of equations (\ref
{non dim current continuity in the Hartmann layer}-\ref{non dim NS in the
hartmann layer}) at the Hartmann walls are :
\begin{eqnarray}
\begin{array}{cc}
\mathbf{u}_{h}\left( \xi =0\right) =0 & w_{h}\left( \xi =0\right) =0
\end{array}
\label{Perpendicular walls conditions} \\
j_{\xi }(\xi =0)=j_{W}
\label{electric condtitions at the parpendicular walls}
\end{eqnarray}
At the edge of the Hartmann layer, the condition of matching with the core
solution implies, for any quantity $g$ (velocity, current density and
pressure) :
\begin{equation}
\underset{\xi \rightarrow +\infty }{\lim }g_{h}=g\left( z=0\right) \equiv
g^{-}  \label{line-up conditions}
\end{equation}
In the case of a free surface at $z=1$ (case $n=1$), this yields :
\begin{equation}
\begin{array}{ccc}
j_{z}(x,y,1)=0, & w(x,y,1)=0, & \partial _{z}\mathbf{u}_{\bot }(x,y,1)=0
\end{array}
.  \label{free surface conditions}
\end{equation}
In the case of a flow between two walls the same condition applies in the
plane of symmetry at $z=1/2$.

We are interested in the limit $Ha\gg 1$ and $N\gg 1$, so that each quantity 
$g$ is developed in term of two small parameters :

\begin{equation}
g=g^{(0)}+g^{\left( 1,0\right) }\dfrac{1}{N}+g^{\left( 0,1\right) }\dfrac{1}{%
Ha}+g^{\left( 1,1\right) }\dfrac{1}{HaN}+...
\label{big areas asymptotic devleoppement}
\end{equation}
In this expansion, the aspect ratio $\lambda $ is supposed fixed, and each
term depends on $\lambda $. The zero order equations in the Hartmann layer
are given by keeping only the right hand terms of (22), (\ref{non dim
current continuity in the Hartmann layer}) and (25). Taking into account the
boundary conditions (26) and the matching conditions (\ref{line-up
conditions}) gives the classical Hartmann layer profile :

\begin{eqnarray}
\begin{array}{cc}
w_{h}^{(0)}=0 & \mathbf{u}_{h}^{(0)}=\mathbf{u}_{\bot }^{-(0)}\left(
1-e^{-\xi }\right) ,
\end{array}
\label{O order Hartmann layer velocities} \\
\begin{array}{cc}
j_{\xi }^{(0)}=j_{W} & \mathbf{j}_{h}^{(0)}=\left( \mathbf{u}_{\bot
}^{-(0)}\times \mathbf{e}_{z}\right) e^{-\xi }
\end{array}
.  \label{O order Hartmann layer current density}
\end{eqnarray}
For the core flow at zero order, (\ref{adim NS in the core}) reduces to 
$\mathbf{j}_{\bot }^{(0)}=0$. The current conservation (\ref{current
continuity core}) then implies $\partial _{z}j_{z}=0,$ so that $%
j_{z}^{(0)}=0 $ if the wall is insulating or if the current $j_{W},$ $%
\partial _{x}j_{W}$ and $\partial _{y}j_{W}$ are much smaller than unity ( $%
j_{W}\ll \lambda \sigma BU$ and $\partial _{x}j_{W}\ll $ $a\sigma BU$ in
physical units)\footnote{%
Notice that these conditions are not achieved at the electrodes where the
current is injected, giving rise to a 3D\ velocity profile, but this effect
will be neglected in forward calculations (section \textbf{4}) as the
surface involved is small in front of the considered domain and the
resulting error on the $z$-average quantities is generally small and
localized.}. Then (\ref{curl of j}) yields $\partial _{z}w^{(0)}=0$ and $%
\partial _{z}\mathbf{u}_{\bot }^{(0)}=0$ so that the flow is 2D in the core.
The pressure $p^{(0)}$ is also 2D, as it results from (\ref{adim NS vertical
in the core}). Matching with the Hartmann layer solution yields :
\begin{equation}
\begin{array}{ccc}
w^{(0)}=0, & \mathbf{u}_{\bot }^{(0)}=\mathbf{u}_{\bot }^{-(0)}\left(
x,y\right) , & \mathbf{\nabla }_{\bot }.\mathbf{u}_{\bot }^{(0)}=0
\end{array}
.  \label{2D core velocity profile}
\end{equation}
This solution corresponds to what we call the ''2D core model'' in section 
\textbf{2.2.1.}

At this point, it should be noticed that the scaling (\ref{Characteristic
values}) overestimates the current density and the resulting electromagnetic
action in the core. The two contributions $\mathbf{\nabla }\phi $ and $%
\mathbf{u\times B}$ in the Ohm's law (\ref{Ohm'slaw}) balance each other so
that the order of magnitude of their sum is lower. The current in the core
and the resulting dynamics for $\mathbf{u}_{\bot }^{(0)}$ is then obtained
at the next order in the expansion (in section \textbf{3.3}.).

\subsection{Recirculating flow in the Hartmann layer.}

Let us now find out the way inertia perturbs at the first order the velocity
profile within the Hartmann layer by introducing the zero order solution in
the left hand side of (22) and (25). Neglecting the left hand term (of order 
$Ha^{-1}$ ) in (\ref{divergence of laplace force in the Hartmann layer}), we
get $\partial _{\xi }\left( \mathbf{j}_{h}\times \mathbf{e}_{z}\right)
=-\partial _{\xi }\mathbf{u}_{h},$ so that $\mathbf{j}_{h}^{(1,0)}\times 
\mathbf{e}_{z}=-\mathbf{u}_{h}^{(1,0)}+\mathbf{j}_{h}^{(1,0)}\left( \xi
=0\right) \times \mathbf{e}_{z},$ and (\ref{non dim NS in the hartmann layer}%
) becomes :

\begin{equation}
\lambda \left( \partial _{t}\mathbf{u}_{h}^{(0)}\mathbf{+u}_{h}^{(0)}.\nabla
_{\bot }.\mathbf{u}_{h}^{(0)}+\nabla _{\bot }p_{h}^{(0)}\right) =\partial
_{\xi \xi }^{2}\mathbf{u}_{h}^{(1,0)}-\mathbf{u}_{h}^{(1,0)}+\mathbf{j}%
_{h}^{(1,0)}\left( \xi =0\right) \times \mathbf{e}_{z}.
\label{1/N Order Ns in the HA layer}
\end{equation}
\bigskip

Therefore, the perturbation $\mathbf{u}_{h}^{(1,0)}$ satisfies a linear
equation in $\xi $ with a source term provided by the zero order solution on
the left hand side. The parallel component of the momentum equation shows
that the pressure is constant along any vertical line at orders $0$, $%
\lambda /N$ and $1/Ha$, so that $p_{h}^{(0)}=p^{-(0)}=p^{(0)}.$ Using the
zero order solution (\ref{O order Hartmann layer velocities}), the no-slip
condition at the wall (\ref{Perpendicular walls conditions}) and matching
with the core flow brings to the expression of $\mathbf{u}_{h}^{\mathbf{(}1,0%
\mathbf{)}}$ :

\begin{multline}
\mathbf{u}_{h}^{(1,0)}=\mathbf{u}^{-(1,0)}\left( 1-e^{-\xi }\right) +\lambda
\left( \dfrac{1}{3}e^{-2\xi }-\dfrac{1}{3}e^{-\xi }+\xi e^{-\xi }\right) 
\mathbf{u}_{\bot }^{-(0)}.\nabla _{\bot }\mathbf{u}_{\bot }^{-(0)}
\label{Inertial velocity profile in the Hartmann layer} \\
+\dfrac{\lambda }{2}\xi e^{-\xi }\partial _{t}\mathbf{u}_{\bot }^{-(0)}.
\end{multline}
The first term corresponds to the classical Hartmann layer associated with
the first order perturbation in the core flow, while the other terms
describe inertial effects.

Indeed, the first order horizontal velocity field is not divergent-free, so
that a vertical flow of order $(HaN)^{-1}$ occurs that can be appraised
thanks to the continuity equation (\ref{non dim continuity equation in the
hartmann layer}) :

\begin{equation}
w_{h}^{(1,1)}=\lambda \mathbf{\nabla }_{\bot }\mathbf{.}\left( \left( 
\mathbf{u}_{\bot }^{-(0)}\mathbf{.\nabla }_{\bot }\right) \mathbf{u}_{\bot
}^{-(0)}\right) \left\{ -\dfrac{5}{6}+\dfrac{2}{3}e^{-\xi }+\xi e^{-\xi }+%
\dfrac{1}{6}e^{-2\xi }\right\} .
\label{vertical velocity in the inertial Hartman layer}
\end{equation}
In the limit $\xi \rightarrow +\infty $ , this vertical flow tends to :

\begin{equation}
w_{h}^{(1,1)}=-\dfrac{5}{6}\lambda \mathbf{\nabla }_{\bot }\mathbf{.}\left(
\left( \mathbf{u}_{\bot }^{-(0)}\mathbf{.\nabla }_{\bot }\right) \mathbf{u}%
_{\bot }^{-(0)}\right) .  \label{inertial core vertical velocity}
\end{equation}

In an axisymmetric configuration, this would describe an Ekman recirculation
(or tea-cup phenomenon). As a matter of fact, depending on whether the
acceleration variation of a fluid particle located at the top of the
Hartmann layer is positive or negative, the latter will be ejected in the
core flow or pumped down to the Hartmann layer. This effect has been
calculated for the classical Ekman layer, in a rotating frame of reference,
by Nanda and Mohanty (1970)\nocite{Nanda70}, whole Loffredo (1986)\nocite
{Loffredo86} extended to MHD\ the classical solution of Von Karman (1921)
for a boundary layer near a rotating plate. The result (\ref{inertial core
vertical velocity}) generalizes such calculations for any bulk velocity
field $\mathbf{u}_{\bot }$.

In the same way, the electric current (\ref{O order Hartmann layer current
density}) closes in a vertical electric current outside the Hartmann layer
with a $z$ current density of order $\lambda /\left( HaN\right) $, obtained
from the current conservation equation (\ref{non dim current continuity in
the Hartmann layer}).

Lastly, the wall friction associated with the velocity profile including
inertia (\ref{Inertial velocity profile in the Hartmann layer}) writes :

\begin{multline}
\mathbf{\tau }^{-}=Ha\partial _{\xi }\dfrac{1}{N}\mathbf{u}%
_{h}^{(1,0)}\left( \xi =0\right)  \label{Inertial Hartmann friction} \\
=\dfrac{Ha}{N}\mathbf{u}_{\bot }^{-(1,0)}+\dfrac{\lambda Ha}{N}\left\{ 
\dfrac{1}{2}\partial _{t}\mathbf{u}_{\mathbf{\bot }}^{-(0)}+\dfrac{2}{3}%
\left( \mathbf{u}_{\mathbf{\bot }}^{-\left( 0\right) }\mathbf{.\nabla }%
_{\bot }\right) \mathbf{u}_{\bot }^{-\left( 0\right) }\right\}
\end{multline}
Once again, the first term corresponds to the classical linear Hartmann
friction associated with the first order perturbation in the core flow,
while the other terms describe the viscous friction associated with inertial
effects.

\subsection{First order perturbation in the core.}

\subsubsection{Recovering the 2D core equation.}

The equation which governs the zero order quantities is derived from the
first order in the expansion. Indeed, the left hand side of (\ref{adim NS in
the core}) can be approximated \ using the zero order velocity

\begin{equation}
\dfrac{\lambda }{N}\left( \partial _{t}\mathbf{u}_{\bot }^{(0)}+\mathbf{u}%
_{\bot }^{(0)}.\mathbf{\nabla }_{\bot }\mathbf{u}_{\bot }^{(0)}+\mathbf{%
\nabla }_{\bot }p^{(0)}\right) =\mathbf{j}_{\bot }\times e_{z},
\label{0-order velocity /current eq in the core}
\end{equation}
so that $\mathbf{j}_{\bot }$ does not depend on $z$, and $j_{z}$ is linear
in $z$ due to the current conservation (\ref{current continuity core}).

As shown in section \textbf{3.1}, $\mathbf{j}_{\bot }^{\left( 0\right) }=0$.
This implies that $\sigma UB$ is not the good order of magnitude for $%
\mathbf{j}_{\bot }$(it is still correct that $\mathbf{u}\times \mathbf{B\sim 
}\sigma UB$ and $-\mathbf{\nabla }\phi \mathbf{\sim }\sigma UB$ but their
sum is of a lower order)$.$ Indeed, a non-zero value of the electric current
density within the core only results from the presence of a non
electromagnetic force in the motion equation (such as inertia). A balance
then sets up between the Lorentz force and the other one and both have to be
of the same order. Looking for the effects of inertia in the core then
requires that the current be of order $\tfrac{\lambda U^{2}}{aB}$. This
value determines the force $\mathbf{j}_{\bot }\times e_{z}$ where the
current density $\mathbf{j}_{\bot }$ has to be fed by the electric current
coming out of the Hartmann layer.

Introducing the zero order current (\ref{O order Hartmann layer current
density}) in the left hand side of the current conservation equation (\ref
{non dim current continuity in the Hartmann layer}) yields the distribution
of vertical current $j_{\xi }^{\left( 0,1\right) }=j_{W}^{\left( 0,1\right)
}-\left( \mathbf{\nabla }_{\bot }\mathbf{\times u}_{\bot }^{-(0)}\right)
\left( 1-e^{-\xi }\right) $\ within the Hartmann layer. Due to the matching
condition (\ref{line-up conditions}), this yields a current $j_{z}^{-\left(
0,1\right) }=j_{W}^{\left( 0,1\right) }-\mathbf{\nabla }_{\bot }\times 
\mathbf{u}^{-(0)}$ at $z=0$ which feeds the core. Since $j_{z}$ is linear in 
$z$, this condition, together with the upper boundary condition (\ref{free
surface conditions}), both determine the vertical current $j_{z}^{\left(
0,1\right) }$, and the corresponding horizontal current $\mathbf{j}_{\bot
}^{\left( 0,1\right) }$ in the core (and the related electromagnetic force $%
\mathbf{j}_{\bot }^{\left( 0,1\right) }\times e_{z}$).
\begin{eqnarray}
j_{z}^{\left( 0,1\right) }&=&\left( 1-nz\right) \left( j_{W}-\dfrac{1}{Ha}%
\mathbf{\nabla }_{\bot }\times \mathbf{u}_{\bot }^{-(0)}\right) ,
\label{vercital current in the core at (0,1)} \\
\mathbf{\nabla }_{\bot }\times \left( \mathbf{j}_{\bot }^{\left( 0,1\right)
}\times \mathbf{e}_{z}\right) &=&\mathbf{\nabla }_{\bot }.\mathbf{j}_{\bot
}^{\left( 0,1\right) }=-\dfrac{1}{Ha}\mathbf{\nabla }_{\bot }\times \mathbf{u%
}_{\bot }^{-(0)}+j_{W}.  \label{Horizontal current divergence (0,1)}
\end{eqnarray}
using (\ref{curl of j}), we can also get $-\mathbf{\nabla }_{\bot }\times 
\mathbf{j}_{\bot }^{\left( 0,1\right) }=\partial _{z}w^{\left( 0,1\right) },$
so that $w$ must be a linear function of $z.$ It vanishes at the free
surface $z=1$ and matches with the Hartmann layer at $z=0$ :
\begin{equation}
w^{\left( 0,1\right) }=w^{-\left( 0,1\right) }\left( 1-nz\right) ,
\label{vertical flow rate of the barrel profile}
\end{equation}
and 
\begin{equation}
\mathbf{\nabla }_{\bot }\mathbf{\times j}_{\bot }^{\left( 0,1\right)
}=-nw^{-\left( 0,1\right) }\mathbf{e}_{z}.  \label{curl of J in the core}
\end{equation}
The vertical component of the velocity $w^{-}$ is given by (\ref{inertial
core vertical velocity}), and scales as $\left( NHa\right) ^{-1}\ll Ha^{-1}.$
Thus $\mathbf{\nabla }_{\bot }\times \left( \mathbf{j}_{\bot }^{\left(
0,1\right) }\times \mathbf{e}_{z}\right) =0,$ and we can write the force in (%
\ref{0-order velocity /current eq in the core}) as : 
\begin{equation}
\mathbf{j}_{\bot }^{\left( 0,1\right) }\times \mathbf{e}_{z}=\mathbf{u}_{0}-n%
\mathbf{u}_{\bot }^{(0)}
\end{equation}
with $\mathbf{u}_{0}=\mathbf{\nabla }_{\bot }\Psi _{0}\times \mathbf{e}_{z}$
and $\Delta _{\bot }\Psi _{0}=-j_{W}$. The order of magnitude of the Lorentz
force in the core is then $\tfrac{\sigma B^{2}U}{\rho Ha}$. As $\left\| 
\mathbf{j}_{\bot dim}\right\| \mathbf{\sim }\tfrac{\lambda U^{2}}{aB}$ , the
effects of inertia are only pertinent if $\tfrac{N}{\lambda Ha}\sim \mathcal{%
O}\left( 1\right) $, so that approximating $\mathbf{j}_{\bot }$ by its
higher order, (\ref{0-order velocity /current eq in the core}) can be
written :
\begin{equation}
\partial _{t}\mathbf{u}_{\bot }^{(0)}+\mathbf{u}_{\bot }^{(0)}.\mathbf{%
\nabla }_{\bot }\mathbf{u}_{\bot }^{(0)}+\mathbf{\nabla }_{\bot }p^{(0)}=%
\dfrac{N}{\lambda Ha}\left( \mathbf{u}_{0}-n\mathbf{u}_{\bot }^{(0)}\right) .
\label{0 Order velocity eq in the core-inertia}
\end{equation}
The cases where $\tfrac{N}{\lambda Ha}$ is not of order one corresponds to
cases where either inertia or Lorentz force are not leading order forces. If 
$\tfrac{N}{\lambda Ha}\ll 1$ , the Lorentz force is not dominant anymore so
that the core flow is not 2D in first approximation : this is the
hydrodynamic case, which is out of our assumptions. In the case $\tfrac{N}{%
\lambda Ha}\gg 1$, inertia is negligible so that the flow is strictly 2D and
adapts instantly to the electromagnetic force. Equation (\ref{0 Order
velocity eq in the core-inertia}) is then still valid in the degenerate form
:
\begin{equation}
\mathbf{u}_{0}=n\mathbf{u}_{\bot }^{(0)}\text{.}  \label{degenerate  2D eq}
\end{equation}
Lastly, using the same method, the effects of viscosity are found to be
relevant if $Ha\sim \lambda ^{2}$. This condition is satisfied in parallel
layers for which $\lambda =Ha^{1/2}$. In the laminar case, inertia is
negligible and assuming that $\mathbf{u}_{\bot }^{\left( 0\right) }$is still
2D (which is a good approximation as shown by figure \textbf{\ref{2D/3D
paralell layer profiles} }and discussed in section \textbf{3.5}) an
equivalent of (\ref{0 Order velocity eq in the core-inertia}) in parallel
layers writes :

\begin{equation}
-\dfrac{\lambda ^{2}}{Ha}\Delta _{\bot }\mathbf{u}_{\bot }^{(0)}=\mathbf{u}%
_{0}-n\mathbf{u}_{\bot }^{(0)},  \label{0-order velocity in the core-viscous}
\end{equation}
Gathering (\ref{0 Order velocity eq in the core-inertia}) and (\ref{0-order
velocity in the core-viscous}) in a single model yields :

\begin{equation}
\dfrac{\lambda }{N}\left( \partial _{t}\mathbf{u}_{\bot }^{(0)}+\mathbf{u}%
_{\bot }^{(0)}.\mathbf{\nabla }_{\bot }\mathbf{u}_{\bot }^{(0)}+\mathbf{%
\nabla }_{\bot }p^{(0)}\right) -\dfrac{\lambda }{Ha}\Delta _{\bot }\mathbf{u}%
_{\bot }^{(0)}=\dfrac{1}{Ha}\left( \mathbf{u}_{0}-n\mathbf{u}_{\bot
}^{(0)}\right) ,  \label{0 Order velocity eq in the core}
\end{equation}

\subsubsection{\protect\bigskip 3D effects in the core : the ''barrel''
effect.}

Let us now investigate the occurrence of 3D effects in the core flow. At
first order, (\ref{adim NS in the core}) takes the general form :

\begin{equation}
\mathbf{F=j}_{\bot }^{\left( 0,1\right) }\mathbf{\times e}_{z},
\label{general action balance in the core}
\end{equation}
where the small quantity $\mathbf{F=}\dfrac{\lambda }{N}\left( \partial _{t}%
\mathbf{u}_{\bot }^{(0)}+\mathbf{u}_{\bot }^{(0)}.\mathbf{\nabla }_{\bot }%
\mathbf{u}_{\bot }^{(0)}+\mathbf{\nabla }_{\bot }p^{(0)}\right) $ depends on
the zero order velocity, which is two-dimensional, so that $\mathbf{j}_{\bot
}\mathbf{\times e}_{z}$ is independent of the vertical coordinate as already
stated. Then the electromagnetic equations (21), and their consequences (\ref
{vertical flow rate of the barrel profile}) and (\ref{curl of J in the core}%
), yield the 3D\ perturbation in velocity. Indeed introducing (\ref{vertical
flow rate of the barrel profile}) and (\ref{curl of J in the core}) in (\ref
{adim Rot of the Ohm's law in the core})(differentiating in $z$ and taking
into account that $\partial _{zz}^{2}j_{z}=-\partial _{z}\left( \mathbf{%
\nabla }_{\bot }.\mathbf{j}_{\bot }\right) =0$) brings to the vertical
dependence of the velocity profile :

\begin{equation}
-\lambda ^{2}\left[ \mathbf{\Delta }_{\bot }\mathbf{F}-\mathbf{\nabla }%
_{\bot }\left( \mathbf{\nabla }_{\bot }.\mathbf{F}\right) \right] =\partial
_{zz}^{2}\mathbf{u}^{\left( 0,1\right) }
\end{equation}
Since the action $\mathbf{F}$ is not dependent on the vertical coordinate,
the response of the flow must exhibit a parabolic velocity profile. Moreover
the free surface condition $\partial _{z}\mathbf{u}_{\bot }=0$ at $z=1$ (or $%
z=1/2$ in the case of two Hartmann walls, $n=2$) yields :

\begin{equation}
\mathbf{u}_{\perp }^{\left( 0,1\right) }(x,y,z)\mathbf{=u}^{-\left(
0,1\right) }(x,y)+\dfrac{1}{2}z\left( z-\dfrac{2}{n}\right) \lambda ^{2}%
\mathcal{L}\mathbf{F(x,y),}  \label{general 3D profile}
\end{equation}
The operator $\mathcal{L}$\ is defined by :

\begin{equation}
\mathcal{L}:\mathbf{F\longmapsto }\mathcal{L}\mathbf{F=-\Delta }_{\bot }%
\mathbf{F+\nabla }_{\bot }\left( \mathbf{\nabla }_{\bot }\mathbf{.F}\right)
\end{equation}

If no flow is injected through the upper or lower boundaries of the core (%
\textit{i.e. }$w^{-}=0$) then the horizontal induced current in the core is
irrotational. The physics leading to this result can be easily understood :
according to (\ref{general action balance in the core}), introducing a 2D\
force (or acceleration) in the core induces a 2D (divergent) horizontal
electric current in the core. To feed the latter, a vertical electric
current has to appear such that $\ j_{z}(x,y,z)-j^{-}(x,y)\sim z$. The
related electric potential is then quadratic : $\phi (x,y,z)\sim \phi
^{-}(x,y)z^{2}.$ As $\mathbf{j}_{\bot }$ is 2D, the Ohm's law $\mathbf{j}%
_{\bot }\mathbf{=}\nabla \phi +\mathbf{u}_{\bot }\times \mathbf{B}$ requires
a quadratic velocity $\mathbf{u}_{\bot }(x,y,z)\sim \mathbf{u}_{\bot
}^{-}(x,y)z^{2}$.

Therefore, adding a 2D\ force not only adds a 2D additional electromagnetic
reaction, but introduces a 3D\ component in the velocity profile. Vortices
do not appear as ''columns'' as described in Sommeria and Moreau (1982)
anymore, but may rather look like ''barrels'' , as the ''cigars'' found by
M\"{u}ck \textit{et Al}.(2000)\nocite{Muck00} thanks to Direct Numerical
Simulations.

Writing explicitly $\mathbf{F}$ in relation (\ref{general 3D profile}) and
using the zero-order evolution equation (\ref{0 Order velocity eq in the
core}), we get :

\begin{equation}
\mathbf{u}_{\perp }\mathbf{=u}_{\perp }^{-\left( 0\right) }-\dfrac{1}{2}%
z\left( z-\dfrac{2}{n}\right) \dfrac{\lambda ^{2}}{Ha}\mathbf{\Delta }_{\bot
}\left( \mathbf{u}_{0}-\mathbf{u}_{\bot }^{-(0)}\right) .
\label{Inertial core profile}
\end{equation}
Notice that the term in $N^{-1}$ is cancelled because of the evolution
equation, so that the resulting perturbation is in $Ha^{-1}$.

This result can be interpreted in terms of the electromagnetic diffusion
time $t_{d}=\tfrac{\lambda ^{2}\rho }{\sigma B^{2}}$ as discussed by
Sommeria and Moreau (1982). Considering the zero-order solution of (\ref
{adim NS in the core}) is equivalent to setting an infinite interaction
parameter and Hartmann number, and thus a zero electromagnetic \ momentum
diffusion time. That means that velocity differences between transverse
planes are instantly damped so that the core flow is 2D. By contrast,
considering a finite diffusion time, the velocity differences are not
completely removed and the parabolic profile appears at first order.

\subsection{Summary of the former developments and commentary.}

Gathering the correction to the 2D\ profile respectively due to the barrel
effect and the recirculating flow occurring in the Hartmann layer yields a
new vertical profile of horizontal velocity. Notice that the full
calculation requires the profiles of $\mathbf{u}_{h}^{(0,1)}=\mathbf{u}%
_{\bot }^{-(0,1)}\left( 1-e^{-\xi }\right) ,$ $\mathbf{u}_{h}^{(1,1)}=%
\mathbf{u}_{\bot }^{-(1,1)}\left( 1-e^{-\xi }\right) $

$+\left( \dfrac{1}{3}e^{-2\xi }-\dfrac{1}{3}e^{-\xi }+\xi e^{-\xi }\right)
\left( \mathbf{u}_{\bot }^{-\left( 0,1\right) }.\mathbf{\nabla }_{\bot }%
\mathbf{u}_{\bot }^{-\left( 0\right) }+\mathbf{u}_{\bot }^{-\left(
0,1\right) }.\mathbf{\nabla }_{\bot }\mathbf{u}_{\bot }^{-\left( 0\right)
}\right) $

$+\dfrac{\xi }{2}e^{-\xi }\partial _{t}\mathbf{u}_{\bot }^{-\left(
0,1\right) },$ $\mathbf{u}_{\bot }^{(0,1)}=$ $\mathbf{u}_{\bot }^{-(0,1)}$%
and $\mathbf{u}_{\bot }^{(1,1)}=\mathbf{u}_{\bot }^{-(0,1)}$, which are
obtained by exactly the same calculations as in sections \textbf{3.1}, 
\textbf{3.2} and \textbf{3.3}. Summing all these terms and using (\ref{big
areas asymptotic devleoppement}) yields the final expressions for the
velocities :

In the Hartmann layer, we have :

\begin{multline}
\mathbf{u}_{h}=\mathbf{u}_{\bot }^{-}\left( 1-e^{-\xi }\right) +\dfrac{%
\lambda }{N}\left( \dfrac{1}{3}e^{-2\xi }-\dfrac{1}{3}e^{-\xi }+\xi e^{-\xi
}\right) \mathbf{u}_{\bot }^{-}.\mathbf{\nabla }_{\bot }\mathbf{u}_{\bot
}^{-}+  \label{Horizontal velocity in the Ha Layer} \\
+\dfrac{\lambda }{N}\dfrac{\xi }{2}e^{-\xi }\partial _{t}\mathbf{u}_{\bot
}^{-}+O\left( \dfrac{\lambda ^{2}}{Ha^{2}}\right) +...
\end{multline}
and

\begin{equation}
w_{h}=\dfrac{\lambda }{HaN}\mathbf{\nabla }_{\bot }\mathbf{.}\left[ \left( 
\mathbf{u}_{\bot }^{-}\mathbf{.\nabla }_{\bot }\right) \mathbf{u}_{\bot }^{-}%
\right] \left\{ -\dfrac{5}{6}+\dfrac{2}{3}e^{-\xi }+\xi e^{-\xi }+\dfrac{1}{6%
}e^{-2\xi }\right\} +...  \label{Vertical velocity in the Ha layer}
\end{equation}
Note that $w_{h}$ induces a vertical velocity component $w=-\dfrac{5}{6}%
\dfrac{\lambda }{HaN}\mathbf{\nabla }_{\bot }\mathbf{.}\left[ \left( \mathbf{%
u}_{\bot }^{-}\mathbf{.\nabla }_{\bot }\right) \mathbf{u}_{\bot }^{-}\right]
(1-nz)$ in the core.

The horizontal velocity $\mathbf{u}_{\bot }(x,y,z,t)$ in the core is given
by (\ref{Inertial core profile}) and it contains no term in $N^{-1}$ :

\begin{multline}
\mathbf{u}_{\bot }(x,y,z,t)=\mathbf{u}_{\bot }^{-}(x,y,t)+\dfrac{\lambda ^{2}%
}{Ha}\dfrac{1}{2}z\left( z-\dfrac{2}{n}\right) \mathbf{\Delta }_{\bot
}\left( \mathbf{u}_{0}-\mathbf{u}_{\bot }^{-}\right)
\label{Horizontal velocity in the core} \\
+O\left( \dfrac{\lambda ^{4}}{Ha^{2}},\dfrac{\lambda ^{3}}{HaN}\right) +...
\end{multline}

The velocity field is therefore determined from the velocity $\mathbf{u}%
_{\bot }^{-}\left( x,y\right) $ close to the wall (but outside the Hartmann
layer). Each order $\mathbf{u}_{_{\bot }}^{-(i,j)}$ of this field $\mathbf{u}%
_{\bot }^{-}$ evolves with time according to an effective 2D equation which
can be obtained at the next order of the expansion. However, it is simpler
to use the average equation (\ref{2D integrate motion equation}), as
performed in next section.

\subsection{A new effective 2D model.}

Two kinds of 3D mechanisms have been pointed out in previous sections : the
recirculating flow in the Hartmann layer, of order $1/N$ and the ''barrel''
effect in the core of order $1/Ha$. Both of them alter the Reynolds tensor
and the upper and lower wall stresses, appearing in (\ref{2D integrate
motion equation})\textbf{. }As\textbf{\ }inertial effects are investigated,
we now restrict the analysis to them and discard the $z$ dependence of the
horizontal velocity in the core ; but in comparison with the 2D\ core model (%
\ref{0-ordrer integrate equation}), vertical velocities are allowed.

Notice that as two different scalings have been used for the Hartmann layer
and the core flow, the vertical average of any quantity $g$ is computed
using $\bar{g}=\int_{0}^{1}gdz+\dfrac{n}{Ha}\int_{0}^{+\infty }g(Ha$ $%
z)-g\left( z=0\right) d\left( Ha\text{ }z\right) .$ With these vertical
velocity profiles, (\ref{Horizontal velocity in the Ha Layer}) in the
Hartmann layer and $\mathbf{u}_{\bot }(x,y,z,t)=\mathbf{u}_{\bot
}^{-}(x,y,t) $ in the core, the averaged velocity $\mathbf{\bar{u}}_{\bot }$
is related to the velocity $\mathbf{u}_{\bot }^{-}$ in the core near the
wall $z=0$ by

\begin{equation}
\mathbf{\bar{u}}=\left( 1-\frac{n}{Ha}\right) \mathbf{u}_{\bot }^{-}+\dfrac{%
n\lambda }{HaN}\left( \dfrac{5}{6}\mathbf{u}_{\bot }^{-}\mathbf{.\nabla }%
_{\bot }+\dfrac{1}{2}\partial _{t}\right) \mathbf{u}_{\bot }^{-},
\label{mean core velocity}
\end{equation}
where $\mathbf{\bar{u}}$ is then a function of $\mathbf{u}_{\bot }^{-}$, as
well as the velocity profiles (\ref{Horizontal velocity in the core}) and (%
\ref{Horizontal velocity in the Ha Layer}). In order to express the
evolution equation (\ref{2D integrate motion equation}) in terms of the
average velocity $\mathbf{\bar{u}}$, which has the advantage of being 2D and
incompressible, (\ref{mean core velocity}) has to be inverted (taking into
account that $\left( HaN\right) ^{-1}\ll 1$ so that $\mathbf{u}_{\bot
}^{-}\simeq \mathbf{\bar{u}}$ for the highest order terms) :

\begin{equation}
\mathbf{u}_{\bot }^{-}=\left( 1+\frac{n}{Ha}\right) \mathbf{\bar{u}}-\dfrac{%
n\lambda }{HaN}\left( \dfrac{5}{6}\mathbf{\bar{u}.\nabla }_{\bot }+\dfrac{1}{%
2}\partial _{t}\right) \mathbf{\bar{u}}.  \label{u- versus u}
\end{equation}

The wall friction $\tau ^{-}=-Ha\partial _{\xi }\mathbf{u}_{h}\left( \xi
=0\right) $\ is obtained from (\ref{Horizontal velocity in the Ha Layer}),

\begin{equation}
\dfrac{1}{Ha^{2}}\tau ^{-}=\dfrac{1}{Ha}\mathbf{u}_{\bot }^{-}+\dfrac{%
\lambda }{HaN}\left[ \dfrac{1}{2}\partial _{t}\mathbf{u}_{\bot }^{-}+\dfrac{2%
}{3}\mathbf{u}_{\bot }^{-}.\mathbf{\nabla }_{\bot }\mathbf{u}_{\bot }^{-}%
\right] .  \label{wallf friction general expression}
\end{equation}
It can be expressed in terms of the variable $\mathbf{\bar{u}}$, using (\ref
{u- versus u}). Including the top wall friction if $n=2,$ this yields the
total wall stress :

\begin{multline}
\dfrac{1}{Ha^{2}}\mathbf{\tau }_{W}=-\dfrac{n}{Ha}\mathbf{\bar{u}}\left( 1+%
\dfrac{n}{Ha}\right)  \label{Wall friction (vectorial)} \\
-\dfrac{n\lambda }{HaN}\left[ \dfrac{1}{2}\partial _{t}\mathbf{\bar{u}}%
\left( 1+\dfrac{n}{Ha}\right) +\mathbf{\bar{u}}.\mathbf{\nabla }_{\bot }%
\mathbf{\bar{u}}\left( \dfrac{2}{3}+\dfrac{11n}{6Ha}\right) \right] .
\end{multline}
Furthermore, the divergence of the Reynolds tensor appearing in (\ref{2D
integrate motion equation}) writes : 
\begin{equation}
\underline{\underline{\mathbf{\nabla }}}_{\bot }.\overline{\mathbf{u}_{\bot
}^{\prime }{}^{t}\mathbf{u}_{\bot }^{\prime }}=\overline{\mathbf{u}_{\bot
}^{\prime }\mathbf{.\nabla }_{\bot }\mathbf{u}_{\bot }^{\prime }}=\frac{n}{%
2Ha}\mathbf{\bar{u}}-\frac{n\lambda }{HaN}\left( \frac{7}{36}\mathcal{D}_{%
\mathbf{\bar{u}}}+\dfrac{1}{8}\partial _{t}\right) \left( \mathbf{\bar{u}%
.\nabla }_{\bot }\right) \mathbf{\bar{u}}
\label{Inertial tensor general expression}
\end{equation}
where the operator $\mathcal{D}_{\mathbf{v}}$ is defined by :

\begin{equation}
\mathcal{D}_{\mathbf{v}}:\mathbf{F\longmapsto }\mathcal{D}_{\mathbf{v}}%
\mathbf{F=}\left( \mathbf{v.\nabla }_{\bot }\right) \mathbf{F+}\left( 
\mathbf{F.\nabla }_{\bot }\right) \mathbf{v=\nabla }_{\bot }\times \left( 
\mathbf{v\times F}\right) -\mathbf{\nabla }_{\bot }.\mathbf{F}
\end{equation}

Writing explicitly the expressions of $\mathbf{\tau }_{W}$ and $\overline{%
\mathbf{u}_{\bot }^{\prime }\mathbf{.\nabla }_{\bot }\mathbf{u}_{\bot
}^{\prime }}$ in (\ref{2D integrate motion equation}) yields an effective
2D\ system of equations for the average velocity $\mathbf{\bar{u}}$. This
equation can be simplified by introducing the new variables 
\begin{eqnarray}
\mathbf{v}&=&\left( 1+7/\left( 6Ha\right) +11/6Ha^{2}\right) \mathbf{\bar{u},}%
\text{ }\mathbf{v}_{0}\mathbf{=}\left( 1+7/\left( 6Ha\right)
+11/6Ha^{2}\right) \mathbf{u}_{0} \\
p^{\prime }&=&\left( 1+7/\left( 6Ha\right) +11/6Ha^{2}\right) p,t^{\prime
}=\left( 1+n/Ha+n^{2}/Ha^{2}\right) ^{-1}t \\
\alpha &=&1+n/Ha
\end{eqnarray}
\begin{equation}
\mathbf{\nabla }_{\bot }.\mathbf{v}=0  \label{Inertial continuity}
\end{equation}

\begin{equation}
\dfrac{\lambda }{N}\left( \dfrac{d\mathbf{v}}{dt^{\prime }}+\mathbf{\nabla }%
_{\bot }\bar{p}^{\prime }\right) =\dfrac{\lambda ^{2}}{Ha^{2}}\mathbf{\Delta 
}_{\bot }\mathbf{v+}\dfrac{1}{Ha}\left( \mathbf{v}_{0}-n\alpha \mathbf{v}%
\right) \mathbf{+}\dfrac{n\lambda ^{2}}{HaN^{2}}\left( \dfrac{7}{36}\mathcal{%
D}_{\mathbf{v}}+\dfrac{1}{8}\partial _{t^{\prime }}\right) \mathbf{v}.%
\mathbf{\nabla }_{\bot }\mathbf{v}
\end{equation}
or in dimensional form (omitting the subscript ()$_{dim}$) :

\begin{equation}
\dfrac{d\mathbf{v}}{dt^{\prime }}+\mathbf{\nabla }_{\bot }\bar{p}^{\prime
}=\nu \mathbf{\Delta }_{\bot }\mathbf{v+}\dfrac{1}{t_{H}}\left( \mathbf{v}%
_{0}-n\alpha \mathbf{v}\right) \mathbf{+}\dfrac{nt_{H}}{Ha^{2}}\left( \dfrac{%
7}{36}\mathcal{D}_{\mathbf{v}}+\dfrac{1}{8}\partial _{t^{\prime }}\right) 
\mathbf{v}.\mathbf{\nabla }_{\bot }\mathbf{v}  \label{New 2D model}
\end{equation}

Notice that it is possible to build a model accounting for both 3D effects
in the core (barrel effect) and inertial effects occurring in the Hartmann
layer. In practice, a complex 2D equation is obtained including seventh
order derivatives terms. Simplicity, which is among the main advantages of
the 2-D model is then lost. In most laboratory experiments, the effects of
inertia are more crucial because they occur for moderate values of $N$
whereas the barrel effect appears for moderate Hartmann numbers ($Ha$ is
much higher than $N$ in usual experimental conditions).

\begin{figure}
\centering
\includegraphics[width=0.7\textwidth]{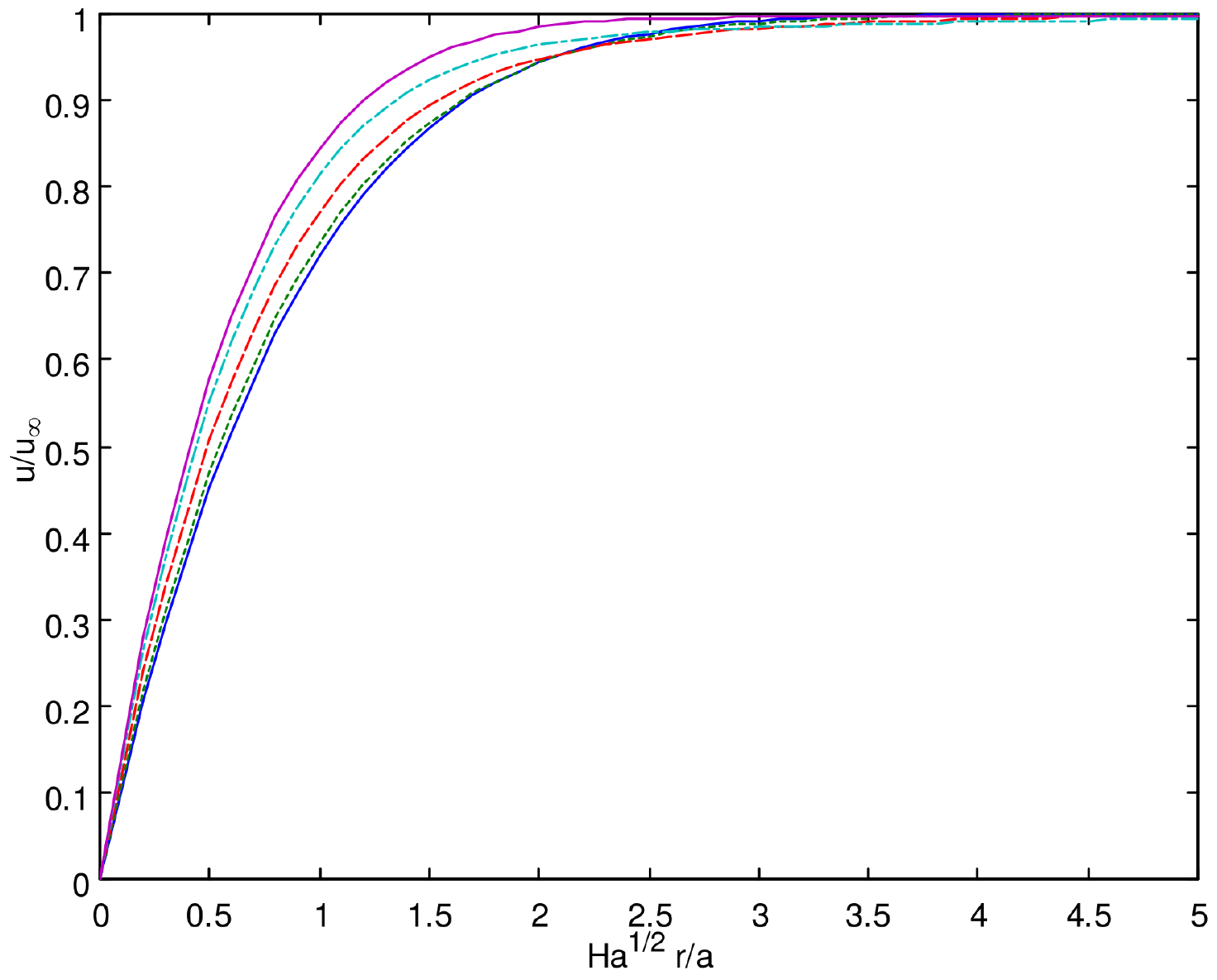}
\caption{\label{Paralelle layer velocity profiles at z=cte.}
Variation with $z$ of the streamwise velocity profile in 3D solutions of
parallel layers (Moreau 1990). inner solid line : $z=0$ , doted line : $%
z=0.4 $ , dashed line : $z=0.7$, dash-dotted line : $z=0.9$, outer solid
line : $z=1$.}
\end{figure}
It is also noticeable that the model built here relies on two assumptions :
the existence of the Hartmann layer and two-dimensionality of the core. The
first one is still rigorously valid in parallel layers as the thickness of
the latter ( $aHa^{-1/2}$ is big in comparison with the Hartmann layer
thickness $aHa^{-1}$). Two-dimensionality is not achieved in parallel layers
but figure \textbf{\ref{Paralelle layer velocity profiles at z=cte.} }shows
that the 3D part of the horizontal velocity field is only $10\%$ of the
velocity. Moreover, this departure is still less relevant since it is
associated to no recirculating velocity, which are the key ingredient by
which the behavior of the flow can be considerably altered. Therefore we
consider that the model can be used in parallel layers, and generates only
small systematic error on the velocities which is not very relevant in
comparison with the correction obtained when accounting for inertial effects
in the Hartmann layer (see examples in section \textbf{4}).

The model (\ref{New 2D model}) has been numerically implemented (work in
preparation). the last term has smoothing properties analogous to a
viscosity. It produces energy decay and spreading of vortices.

\section{Case of axisymmetric flows.}

This section is devoted to the implementation of the former model on simple
axisymmetric flows, which allow explicit calculation and therefore an easy
comparison with the MATUR experiment (Alboussi\`{e}re \textit{et al. }1999 )
and to isolated vortices of Sommeria (1988). For steady axisymmetric flows,
the general expression (\ref{New 2D model}) with non dimensional polar
coordinates, using the previous set of characteristic values (\ref
{Characteristic values}) is strongly simplified as $\partial _{\theta }=0,$ $%
\partial _{t}=0,$ $\bar{u}_{r}=0$. Its azimuthal component yields :

\begin{equation}
\dfrac{7}{36}\dfrac{n\lambda }{HaN^{2}}\dfrac{1}{r^{2}}\partial _{r}\left( r%
\bar{v}_{\theta }^{3}\right) =\dfrac{\lambda }{Ha^{2}}\frac{1}{r^{2}}%
\partial _{r}\left( r^{3}\partial _{r}\dfrac{v_{\theta }}{r}\right) +\dfrac{1%
}{\lambda Ha}\left( v_{0\theta }-n\alpha v_{\theta }\right) .
\label{2D axi equation motion}
\end{equation}

\subsection{Axisymmetric parallel layers.}

We consider here the case of a flow bounded by a vertical cylindrical wall,
a circle of radius $R$ in the 2-D average plane. We seek for the non linear
3-D effects in the boundary layer arising along this wall.\textit{\ }It is
natural to place the frame origin at the center of the circle. Thus, if $R$
is large enough, then in the vicinity of the wall it will be quite justified
to assume that $1/r\approx 1/R<<\partial _{r}$, so that terms which are of
order $1/R^{2}$ are negligible, which leaves equation (\ref{2D axi equation
motion}) under the form :

\begin{equation}
\dfrac{7}{36}\dfrac{Ha}{N^{2}}\dfrac{n}{R}\partial _{r}v_{\theta
}^{3}=\partial _{rr}^{2}v_{\theta }+\dfrac{Ha}{\lambda ^{2}}\left[ -n\alpha
v_{\theta }+v_{0\theta }\right] .  \label{axi para layer motion equation}
\end{equation}
In the case of a concave parallel boundary layer, the following variables
are relevant :

\begin{equation}
y=\left( R-r\right) \sqrt{n\alpha Ha}\text{ and }v_{\theta }=\dfrac{%
v_{0\theta }}{n\alpha }\tilde{v}_{\theta };
\end{equation}

they transform (\ref{axi para layer motion equation}) and the corresponding
boundary conditions in (where $C=\dfrac{7}{36}\dfrac{n^{-3/2}}{\alpha ^{5/2}}%
\dfrac{\sqrt{Ha}}{N}\dfrac{a}{R}$) :

\begin{equation}
\begin{array}{c}
-C\partial _{y}\tilde{v}_{\theta }^{3}=\partial _{yy}^{2}\tilde{v}_{\theta
}+1-\tilde{v}_{\theta }, \\ 
\underset{y\rightarrow +\infty }{\lim }\tilde{v}_{\theta }=1, \\ 
\tilde{v}_{\theta }\left( y=0\right) =0,
\end{array}
\label{adim polar eq for big R}
\end{equation}

\begin{figure}
\centering
\includegraphics[width=0.7\textwidth]{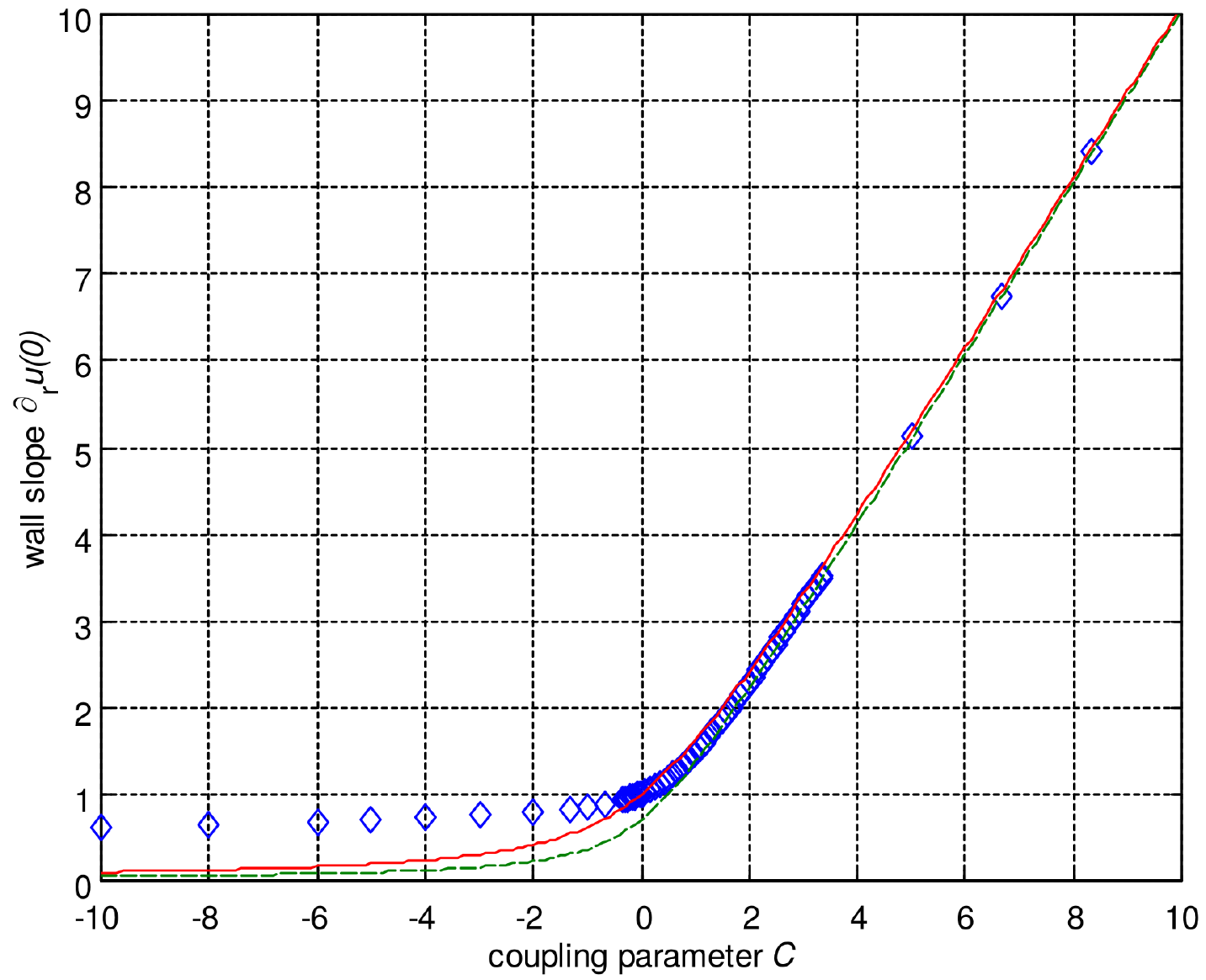}
\caption{\label{curve u'(C)} Velocity profile slope at the wall in
function of the coupling number: \textbf{boxes }: \textbf{\ } boxes\textbf{%
\ }: numerical simulation of equation (\ref{adim polar eq for big R}), solid
line : model (\ref{wall stress in function of C}), dashed line : same as (%
\ref{wall stress in function of C}) when the velocity profile (\ref{velocity
fall}) in the layer is replaced by a straight line.}
\end{figure}

The alternative case of a convex boundary layer, such as the one that would
arise along the outside of a circular cylinder, could be achieved by just
changing the sign of the non dimensional constant $C$. This constant
represents the strength of the inertial transport compared to viscous
dissipation and electric forcing. It is indeed expected to change the
traditional boundary layer profile and the wall friction accordingly. It is
quite relevant since it points out the dissipative role of the boundary
layer which allows to assess the loss of global quantities such as energy or
angular momentum. Therefore numerical computation has been performed that
gives $\partial _{y}\tilde{v}_{\theta }\left( 0\right) $ for a wide range of
values of $C.$ A shooting method featuring a Runge-Kutta algorithm provides
the points plotted in figure \textbf{\ref{curve u'(C)}.}

An analytical approximation provides a reliable description for large values
of $C$. Indeed, (\ref{adim polar eq for big R})\textbf{\ }can be integrated
over $\left[ 0,+\infty \right[ $ to give :
\begin{equation}
\partial _{y}\tilde{v}_{\theta }\left( 0\right) +C+\int_{0}^{+\infty
}\left( \tilde{v}_{\theta }-1\right) dy=0.
\end{equation}
In boundary layers, the velocity fall is strongly concentrated in the
vicinity of the wall, which suggests to replace the profile roughly by an
exponential with $\partial _{y}\tilde{v}_{\theta }\left( 0\right) $ as wall
slope, 
\begin{equation}
\tilde{v}_{\theta }\left( y\right) \simeq 1-\exp \left( -\partial _{y}\tilde{%
v}_{\theta }\left( 0\right) y\right) \text{,}  \label{velocity fall}
\end{equation}
so that

\begin{equation}
\int_{0}^{+\infty }\left( \tilde{v}_{\theta }-1\right) dy=-\dfrac{1}{%
\partial _{y}\tilde{v}_{\theta }\left( 0\right) },
\label{approximation of the profile}
\end{equation}
which brings to the approximate relation :

\begin{equation}
\partial _{y}\tilde{v}_{\theta }\left( 0\right) =\dfrac{C+\sqrt{C^{2}+4}}{2},
\label{wall stress in function of C}
\end{equation}
the asymptotic behavior of which gives a satisfactory fit to numerical
results (see figure \ref{curve u'(C)}).
\begin{eqnarray}
\partial _{y}\tilde{v}_{\theta }\left( 0\right) &\underset{C\rightarrow +\infty }{\sim }&C+O\left( \dfrac{1}{C}\right) \text{for a concave wall,}
\label{u'(0) asymptotic values} \\
\partial _{y}\tilde{v}_{\theta }\left( 0\right) &\sim& -\dfrac{1}{C}+O\left( 
\dfrac{1}{C^{2}}\right) \text{ for a convex wall.}
\end{eqnarray}
In the case of a concave wall, the typical thickness of the parallel layer
is shrunk by the non-linear angular momentum transfer, which feeds wall
dissipation, giving rise to a different kind of boundary layer of typical
non-dimensional thickness $\tfrac{1}{C}$ or $\tfrac{36}{7}\tfrac{N}{Ha}%
Rn^{3/2}$ in physical units. It should be mentioned that this kind of layer
may not be compared to the one resulting from a balance between inertial and
electromagnetic effects (of typical thickness $aN^{-1/3}$) as our parallel
layer does not result from such a balance : it is a classical parallel layer
in which inertial effects driven by the Hartmann layer are taken in account,
which is very different.

This mechanism can be understood as an Ekman pumping whose meridian
recirculation induces an angular momentum flux toward the wall corresponding
to the first term in (\ref{2D axi equation motion}). The radial velocity can
be estimated using the 3D continuity equation (\ref{adim continuity in the
core}) in the core where it reduces to $u_{r}\simeq w^{-}=\tfrac{5}{6}\tfrac{%
\lambda }{HaN}\tfrac{\tilde{v}_{\theta }^{2}}{R}$ (in non dimensional form,
using the initial set of characteristic values(\ref{Characteristic values}%
)). The boundary layer then results from the balance between transport,
forcing and viscous dissipation. When $C$ is large enough, forcing vanishes
from the balance and the boundary layer exclusively dissipates the
transported angular momentum. If the wall is convex ($C<0$), the momentum
flux is reversed, and the boundary layer tends to widen. Figure \textbf{\ref
{curve u'(C)}} shows that the analytical curve (\ref{wall stress in function
of C}) is not pertinent for negative values of $C$ anymore. This is quite
natural as it is justified for a thin boundary layer. Indeed, one can expect
that the larger the latter, the more determinant the shape of the profile is
for the computation of the velocity loss.
\subsection{Consequences on the global angular momentum - the MAgnetic
TURbulence (MATUR) experiment.}
\begin{figure}
\centering
\includegraphics[width=0.7\textwidth]{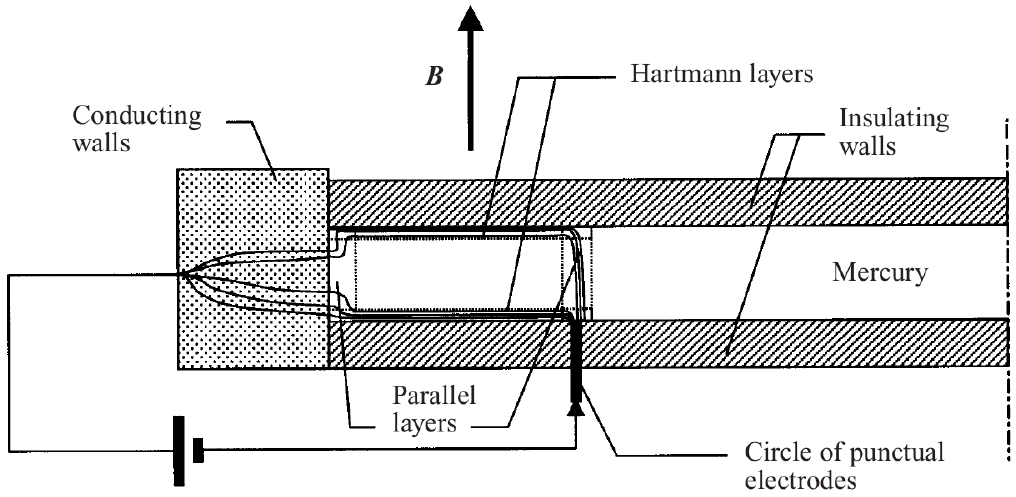}
\caption{\label{Matur scheme} Radial section of Matur experimental
setup.}
\end{figure}
The results of the previous subsection are now compared with experimental
results obtained on the device MATUR. The latter is a cylindric container
(diameter $0.2m$) with electrically insulating bottom and conducting
vertical walls (figure \ref{Matur scheme}). Electric current is injected at
the bottom through a large number of point-eclectrodes regularly spread
along a circle whose center is on the axis of the cylinder. It is filled
with mercury ($1cm$ depth) and the whole device is plunged in a vertical
magnetic field. The injected current leaves the fluid through the vertical
wall inducing radial electric current lines and gives rise to an azimuthal
action on the fluid included in the annulus between the electrode circle and
the outer wall. The injected current $j_{W}$ can be considered as a Dirac
delta function, centered at the injection radius $r_{e}$, with integral
equal to the injected current $I$ : $j_{W}=I/\left( 2\pi r_{e}\right) \delta
\left( r-r_{e}\right) .$ The corresponding \ forcing is azimuthal and given
from the solution of (\ref{current}), which yields : 
\begin{equation}
\mathbf{v}_{0}\simeq \mathbf{\bar{u}}_{0}=-\dfrac{B}{\rho a}\dfrac{I}{2\pi r}%
t_{H}.\mathbf{e}_{\theta }.  \label{Matur forcing}
\end{equation}
This annulus of fluid then rotates and gives rise to a concave parallel
layer along the outer wall. The upper surface of mercury may be either free
or not. But if free, oxidation of mercury makes the upper surface rigid so
that a Hartmann layer takes place at the top anyway. Therefore two Hartmann
layers (at the top and the bottom) have to be considered ($n=2$ ). A more
exhaustive description of the experimental device and results can be found
in Alboussi\`{e}re \textit{et al.} (1999).

\begin{figure}
\centering
\includegraphics[width=0.7\textwidth]{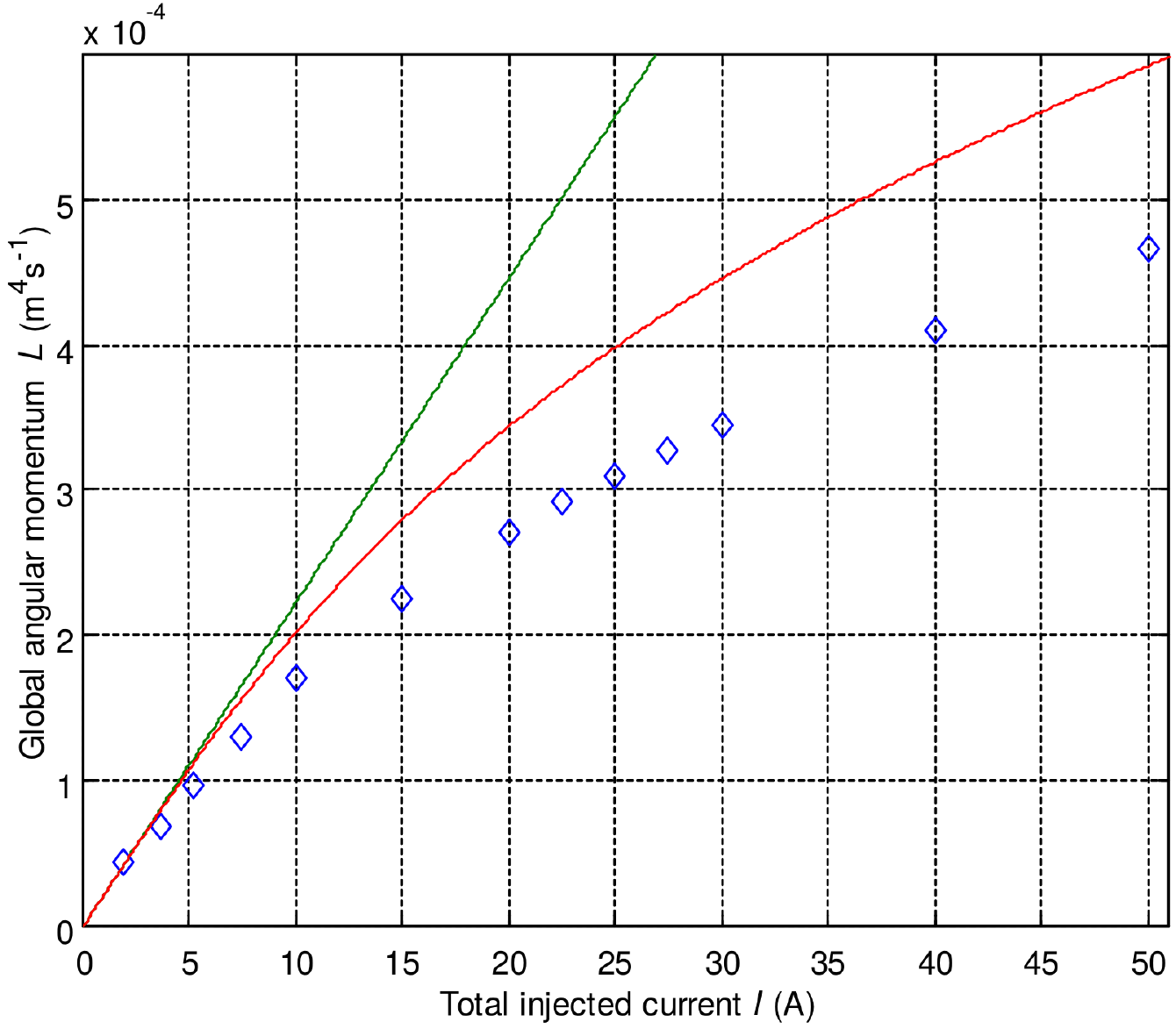}
\caption{\label{Matur Béné.}Global Angular momentum in the Matur
experimental setup versus total injected electric current for $B=0.17T$ and $%
r_{e}=93mm$. Dots : experimental measurements, Solid line : theoretical
curve for obtained from (\ref{I-L relation}).}
\end{figure}

The geometry of the fluid motion suggests that an Ekman recirculation
occurs, rising up a radial flow toward the parallel wall side layer. One can
expect the angular momentum decrease significantly there, altering the
behavior of the layer. A good global description of this effect is provided
by the balance of the total angular momentum $L=\int r\bar{u}_{\theta }d^{2}%
\mathbf{r}$. The equation for $L$ can be derived by integration over the
whole domain of (\ref{New 2D model}) after multiplication by $r$ (assuming $%
\mathbf{v\simeq \bar{u}}$) :

\begin{equation}
\dfrac{dL}{dt}=F-S-\dfrac{2L}{t_{H}},  \label{Global L equation}
\end{equation}
where the global electric forcing $F$ and the viscous dissipation at the
wall side layer $S$ take the form :

\begin{eqnarray}
F& =\dfrac{IB}{2\rho a}\left( R^{2}-r_{e}^{2}\right) ,
\label{Global injected current expression} \\
S& =2\pi R^{2}\nu \partial _{y}u_{\theta }\rfloor _{wall}
\label{dissipation at the wall}
\end{eqnarray}
At small forcing, the parallel layer thickness is of order $aHa^{-1/2},$ so
the corresponding viscous effect on the angular momentum is negligible in
comparison with the Hartmann friction (in a ratio of order $Ha^{1/2}$).
Therefore $S$ can be neglected in (\ref{Global L equation}) and 
\begin{equation}
F=2\dfrac{L}{t_{H}}  \label{Linear momentum equation}
\end{equation}
in steady regime. This corresponds to the linear behavior of $L$ versus the
forcing current $I$ for moderate $I$ ($I\lesssim 7A$ see figure \textbf{\ref
{Matur Béné.}}). Notice that the velocity near the wall is then derived from
the recirculation $\Gamma ,$ by $U=\Gamma /\left( 2\pi R\right) $ and it
coincides with (\ref{Matur forcing}) at $r=R,$ $U=\mathbf{\bar{u}}_{0}\left(
R\right) .$ Comparing $U$ with $L,$ given from the forcing $F$ by (\ref
{Linear momentum equation}), gives 
\begin{equation}
L=\pi R\left( R^{2}-r_{e}^{2}\right) U  \label{U-L}
\end{equation}

We observe that the velocity profile remains unchanged even for large
currents, so we can use (\ref{U-L}) to express the velocity near the wall as
a function of $L.$ Introducing this velocity $U$ in the boundary layer model
of section \textbf{4.1}, we can deduce the wall stress $S$. We have found
that the asymptotic expression (\ref{u'(0) asymptotic values}) is valid for
the considered experimental conditions, allowing a simple expression of $%
\partial _{y}u_{\theta }\rfloor _{wall}$ in (\ref{dissipation at the wall}).
It is then possible to assess every terms in (\ref{Global L equation})%
\textbf{\ }which provides a relation between the injected electrical current
and the global angular momentum, which can be compared to experimental
results :

\begin{equation}
I=4\dfrac{\sqrt{\sigma \rho \nu }}{R^{2}-r_{e}^{2}}L+\dfrac{7}{18}\dfrac{%
\sqrt{\nu \rho ^{5}\sigma ^{-3}}L^{3}}{\left( R^{2}-r_{e}^{2}\right) ^{3}\pi
^{2}R^{2}B^{4}\alpha ^{3}}.  \label{I-L relation}
\end{equation}

\begin{figure}
\centering
\includegraphics[width=0.7\textwidth]{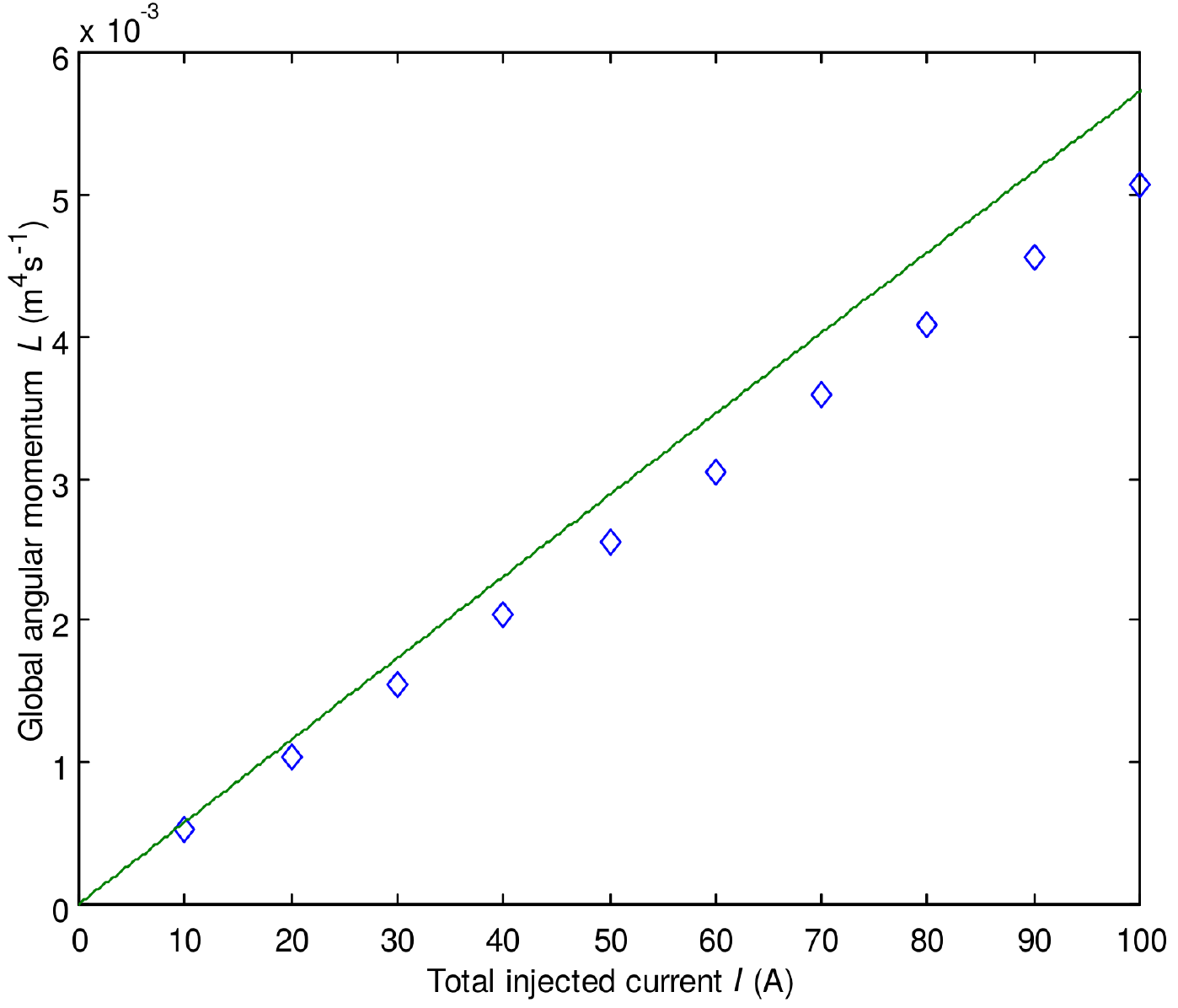}
\caption{\label{Matur B=2T} Global Angular momentum in the Matur
experimental setup versus total injected electric current for $B=2T$ and $%
r_{e}=54mm$. Dots : experimental measurements, Solid line : theoretical
curve obtained from (\ref{I-L relation}).}
\end{figure}

Figures \textbf{\ref{Matur Béné.}} and \textbf{\ref{Matur B=2T}},\textbf{\ }%
show experimental measurements of the global angular momentum and
theoretical curves. Our model provides a reasonable prediction of the
experimental results. This comparison must be put in perspective as MATUR is
a very complex device where a wide variety of phenomena occurs. In
particular, big vortices are present and break the axisymmetry : firstly,
they interact with each other, giving rise to thin shear layers where
dissipation occurs, and secondly they interact with the walls, inducing
separations in the wall side layers. Furthermore, the Hartmann layer may
become turbulent which the present theory does not take in account. Indeed,
one can refer to the heuristic criterion established by Hua and\ Lykoudis%
\textit{\ (1974) }\nocite{Hua74}\textit{\ }which states that in rectangular
ducts, considerable turbulent fluctuations are observed in the vicinity
Hartmann layer for values of $\func{Re}/Ha$ above 250. For $B\geq 0.8T$, the
smallest values of this parameter are about $500$. For all these reasons, it
is natural that our model predicts a dissipation smaller than observed in
the experiment. A numerical simulation of (\ref{New 2D model}) may be able
to take unsteadiness into account and to provide better results.

At higher field (figure \ref{Matur B=2T}\textbf{)}, the saturation has
disappeared from experimental measurements, which are then closer to the
linear theory curve. This is quite natural as the non linear effects are
proportional to $Ha^{-3}$ , which dramatically falls in for increased values
of $B.$ Though the experimental points fit a straight line, the latter has
not exactly the same slope as the one predicted by the linear theory which
is linearly dependent on $1/t_{H}$ . Once again, additional phenomena have
to be invoked. Actually, the bottom of the experimental device contains many
conducting electrodes in which electric current may pass : in these areas,
the damping may be significantly increased, leading to a reduction of the
damping time ''felt'' by the global angular momentum. This latter phenomenon
is certainly responsible for a systematic departure between theory and
experiment.
\subsection{Isolated vortices aroused by a point-electrode. Experimental
comparison.}

The present subsection is devoted to the improvement of the 2D model of
isolated vortices exposed in section \textbf{2.2.3}. taking in account Ekman
recirculation. Indeed, the Sommeria experiments \textbf{(}1988)\textbf{\ }%
clearly show that the core of an isolated vortex tends to widen when the
injected current is strong. As an Ekman secondary flow is highly suspected
of being responsible for this phenomenon, the axisymmetric equation of
motion provides a good analytical model for it.

Let us then consider a configuration similar to the one described in
paragraph \textbf{2.2.3} in which the electric current is injected through a
cylindrical electrode of radius (and an upper free surface so that $n=1$) at
the center of the vortex, introducing a no-slip condition at this point. We
suppose that the forcing satisfies (\ref{pointnelectrode circulation}). The
motion equation (\ref{2D axi equation motion})\ has to be rescaled using the
scalings of section \textbf{2.2.3 }$\lambda =\sqrt{Ha}$ and $U=$ $\dfrac{%
\Gamma }{a}\sqrt{Ha}$. We assume $\mathbf{u}_{0}\simeq \mathbf{v}_{0}$ since 
$Ha\gg 1$, which leaves the non dimensional equation of motion under the
form :

\begin{equation}
C_{t}\dfrac{1}{r^{2}}\partial _{r}\left( rv_{\theta }^{3}\right) =\dfrac{1}{%
r^{2}}\partial _{r}\left[ r^{3}\partial _{r}\left( \dfrac{v_{\theta }}{r}%
\right) \right] -v_{\theta }+\dfrac{1}{r},  \label{non lin tourb adim eq}
\end{equation}
with the corresponding boundary conditions (where $C_{t}=\dfrac{7}{36}\dfrac{%
\Gamma }{a^{2}}\dfrac{1}{N^{2}}$) :

\begin{equation}
\begin{array}{c}
v_{\theta }\left( b\right) =0 \\ 
\text{and}\underset{r\rightarrow +\infty }{\lim }v_{\theta }=0,
\end{array}
\label{non lin tourb boundary conditions}
\end{equation}

One can also express the non dimensional number $C_{t}$ in function of the
local interaction parameter $N_{c}$ introduced by Sommeria\textit{\ }(1988) :

\begin{equation}
\begin{array}{cc}
N_{c}=\dfrac{\sigma B^{2}}{\rho }\dfrac{a^{2}}{\Gamma Ha}, & C_{t}=\dfrac{7}{%
36}\dfrac{1}{N_{c}^{2}}.
\end{array}
\label{coupling number vs interaction parameter}
\end{equation}

The latter result, shows that the local interaction parameter is the
relevant non-dimensional number which controls the radial profile of
azimuthal velocity of the vortex.

\begin{figure}
\centering
\includegraphics[width=0.7\textwidth]{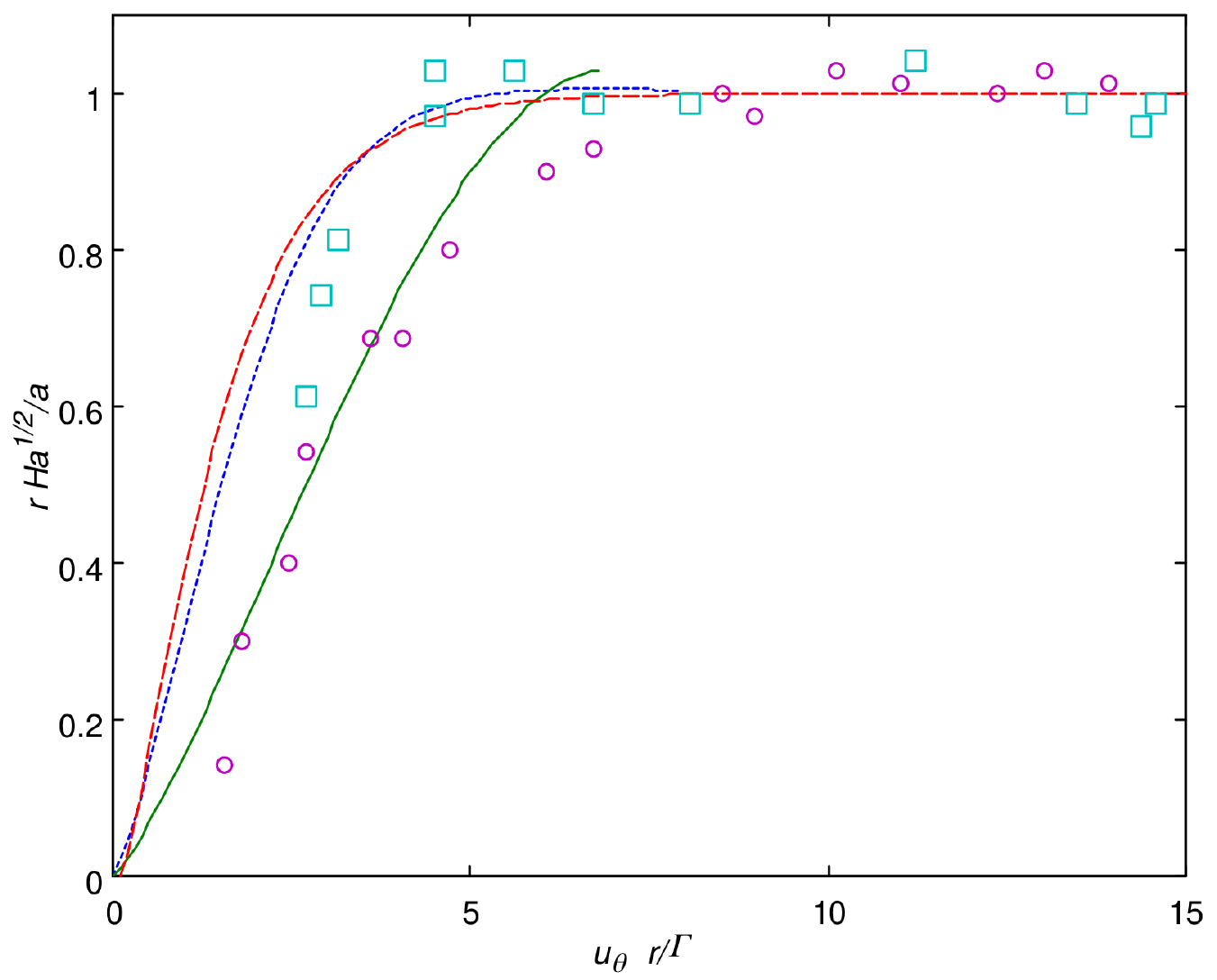}
\caption{\label{Vortex radial profile of angular momentum}Vortex
radial profile of angular momentum for $B=0.5T$ : boxes : experimental
measurements for injected current $I=0.05A,$ circles : experimental
measurements for $I=0.2A,$ full line : analytical profile without non linear
effects, semi dotted line : numerical profile for $I=0.05A$, dotted line :
numerical profile for $I=0.2A$. Note that numerical precision problems don't
allow to get to profiles for any values of r.}
\end{figure}

For an electrode radius $b=0.1,$ the rod is ten times thinner than the
typical parallel layer scale so that this case may be relevantly compared to
the experimental case. Indeed, reducing the electrode diameter when the
latter is significantly smaller than $1$ do not have any relevant effect on
the result. The solutions have been numerically computed thanks to a
shooting method featuring a Runge-Kutta algorithm. The radial profile of
angular momentum has been processed out from the result . We have computed
it for $B=0.5T$ and for two different injected currents ($I=50mA$ and $%
I=200mA$ respectively corresponding to $C_{t}=8.85$ and $C_{t}=141.65$ ) ;
these cases are thus highly non-linear and one can expect the recirculating
flow to be significant. The profiles are reported in figure \textbf{\ref
{Vortex radial profile of angular momentum}} and compared with the
experimental results obtained by Sommeria (1988). The radial velocity can be
estimated from the continuity equation (\ref{adim continuity in the core}) $%
u_{r}=\tfrac{5}{6}\tfrac{\lambda }{HaN}\tfrac{\tilde{v}_{\theta }^{2}}{r}$
in non dimensional form, using (\ref{Characteristic values}).

\begin{figure}
\centering
\includegraphics[width=0.7\textwidth]{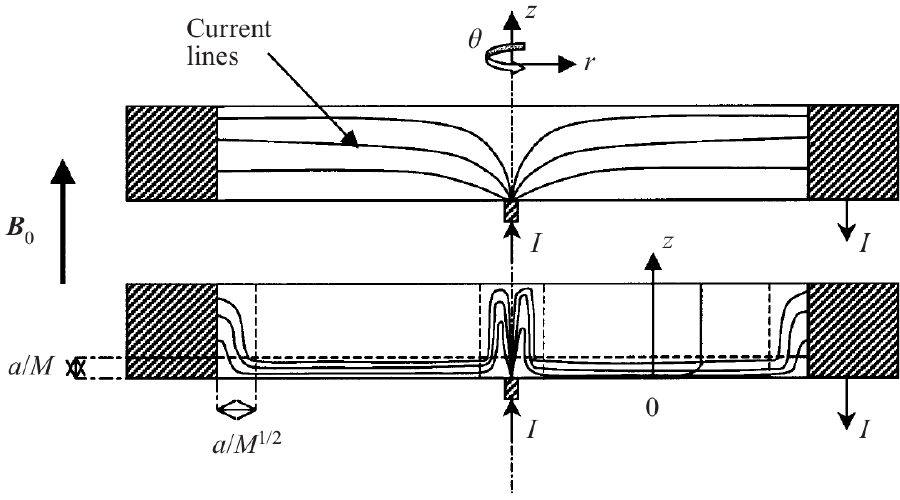}
\caption{\label{The electric current streamlines} The electric
current streamlines (a) without Magnetic field (b) in a strong magnetic
field. The Hartmann layer, the outer layer parallel to the field and the
vortex core are represented, as well as a vertical velocity profile.}
\end{figure}

The experimental device used by Sommeria is similar to the experiment MATUR
except that the electrical current is injected through a single central
electrode and the upper surface is free (see figure \ref{Somméria's Vortex
study experimental device} and \ref{The electric current streamlines}\textbf{%
.}). The velocity measurements are obtained thanks to a visualization
technique including streak photos of particles in the fluid. The numerical
simulations performed using our non linear model gives a good agreement with
the experimental results : it turns out that the vortex core actually
broadens for higher values of the electric current, \textit{i.e. }for
highest values of $C_{t}$). This is due to a radial flow resulting from
inertial effects. Indeed, in axisymmetric configuration, the vertical flow $%
w^{-}$ (\ref{Vertical velocity in the Ha layer}) is proportional to $%
1/r\partial _{r}\left( u_{\theta }^{2}/r\right) $ so that a strong flow rate
from the Hartmann layer occurs at the center of the vortex. This flow softly
closes at large $r$ , which is analogous to the traditional Ekman pumping.

To quantify this phenomenon of spreading vortices, we have plotted the
radius $R_{v}$ of the vortex obtained from the numerical simulations of (\ref
{non lin tourb adim eq}) versus the value of the core related interaction
parameter $N_{c}$.The results are plotted in figure \textbf{\ref{Vortex core
radius}} and it appears that $R_{v}\sim N_{c}^{-1}$. This scaling law is in
agreement with experimental measurements of Sommeria (1988). However, a
quantitative comparison of the prefactor is difficult because the
experimental results are derived from electric potential measurements which
are sensitive to the singularity at the electrode.
\begin{figure}
\centering
\includegraphics[width=0.7\textwidth]{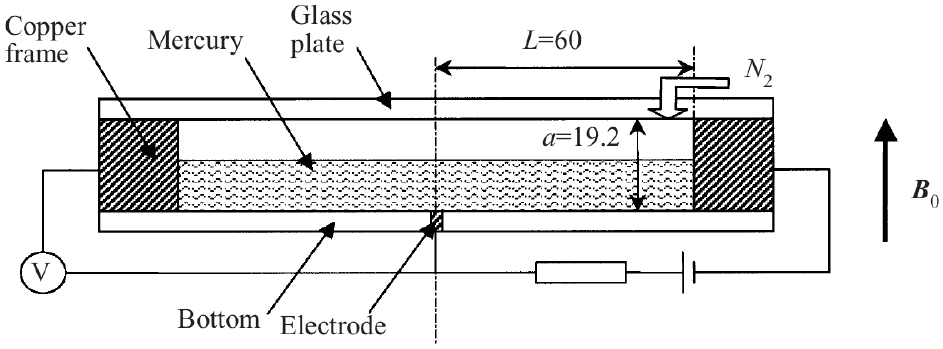}
\caption{\label{Somméria's Vortex study experimental device}
Experimental device of Sommeria's vortex study : cross section of the
circular tank with a shematic representation of the current supply and
device for potential measurement. Dimensions are indicated in mm.}
\end{figure}
\begin{figure}
\centering
\includegraphics[width=0.7\textwidth]{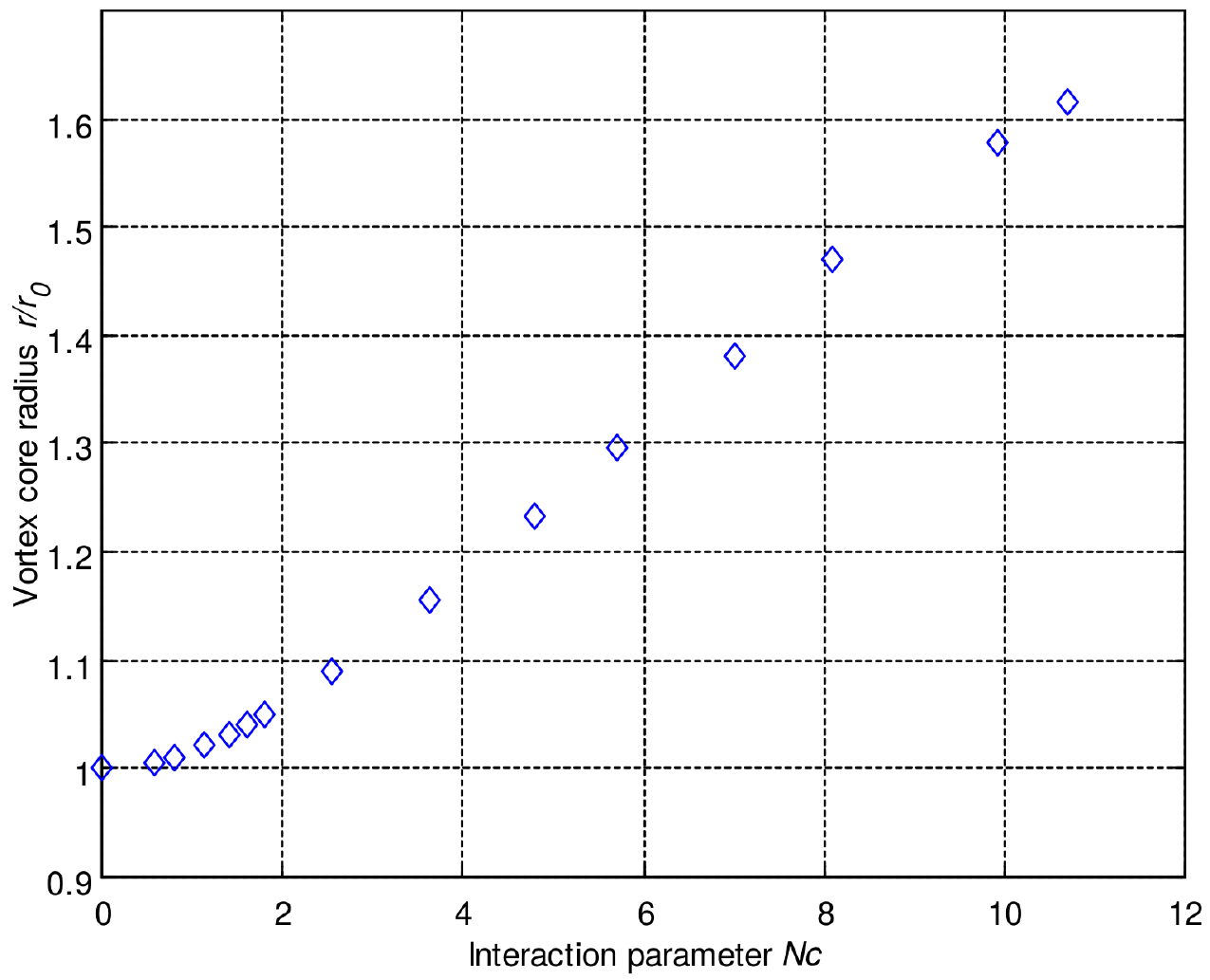}
\caption{\label{Vortex core radius} Vortex core radius normalized by
its value without non-linear effect $r_{0}$(corresponding to very low
injected electric current) versus the inverted core interaction parameter.
The results are obtained from numerical simulations of equation (\ref{non
lin tourb adim eq}).}
\end{figure}
\section{Conclusion.}

Our analysis applies for flows in ducts with transverse uniform magnetic
field, a standard configuration of interest in various MHD problems. These
flows often involve complex 3D velocity fields, with both transverse
structures and vertical variations in the thin Hartmann boundary layer. The
latter is very difficult to resolve numerically at high Hartmann number, due
to the high spatial resolution required. Our effective 2D model provides
thus a great \ simplification.

This model has been derived by a systematic expansion in terms of the two
small parameters $1/N$ and $1/Ha$ providing \ a good understanding of its
range of validity. At zero order, we recover the 2D core model of Sommeria
\& Moreau (1982) which is already a good approximation even in parallel
layers near lateral walls or around a central electrode scaling like $%
Ha^{-1/2}$ (as seen in section \textbf{2}).

The expansion is valid for sufficiently large transverse scales $\lambda $.
In principle $\lambda <Ha^{-1/2}$ has to be satisfied, but the zero order
solution turns out to provide good results even in parallel layers, of
thickness of order $Ha^{-1/2}$. Perturbations at the scale of order $Ha^{-1}$%
can arise in the Hartmann layer, when it becomes unstable. This is
experimentally observed for $\dfrac{N}{Ha}>250$. Such a small scale effect
is not captured by our expansion.

A first correction to the 2D\ core model occurs as weakly three-dimensional
velocity profiles parabolic in $z$ at first order. This effect can be
interpreted as the consequence of the finite diffusion time of momentum by
electromagnetic effects. This diffusion leads to complete two-dimensionality
only in the limit of very large magnetic field ($Ha\rightarrow \infty $).
Vortices look like ''barrels'' instead of columns. We however find that this
essentially linear effect has little influence on the global dynamics,
involving $z$-averaged quantities.

The second perturbation corresponds to Ekman recirculation effects within
the Hartmann boundary layers. This recirculation transports momentum, which
significantly modifies the dynamics of the $z$-averaged velocity. These
recirculating effects can also have interesting consequences for the
transport of heat or chemicals away from the Hartmann layers.

Analytical solutions of our effective 2D model in axisymmetric
configurations appear in reasonable agreement with laboratory experiments.
The model explains the additional dissipation of angular momentum due to
radial transport by recirculation. For the experiments of Sommeria (1988),
it explains the spreading of vortex core and fits the experimental law in $%
N^{-1}.$

Finally, it is noteworthy that recirculation effects lead to new scaling
laws for side layers along concave or convex walls parallel to the magnetic
field. Along a convex wall, the side layer is widened according to equation (%
\ref{adim polar eq for big R}) whose numerical solution is plotted in figure 
\ref{curve u'(C)}. Along a concave wall, on the contrary, it becomes thinner
and the scaling law is in $NHa^{-1}$.

\begin{center}
{\Large References}
\end{center}

T.ALBOUSSIERE, V.USPENSKI \& R.MOREAU 1999 Quasi-2D MHD Turbulent Shear
Layers Experimental thermal and Fluid Science 20 pp19-24.

A.ALEMANY, R.MOREAU, P.SULEM \& U.FRISCH 1979 Influence of an External
Magnetic Field on Homogeneous MHD Turbulence \textit{Journal de Mecanique} 
\textbf{18(2) pp }277-313.

L. B\"{U}HLER 1996 Instabilities in Quasi Two-Dimensional
Magnetohydrodynamic Flows\textit{\ J. Fluid. Mech }\textbf{326 }125-150.

P.A.DAVIDSON 1997 The Role of Angular Momentum in the Magnetic Damping of
Turbulence \textit{J. Fluid. Mech. }\textbf{336 }123-150.

H.M.HUA \& P.H.LYKOUDIS\ 1974 Turbulent Measurements in Magneto-Fluid
Mechanics Channel \textit{Nucl. Sci. Eng.}\textbf{45} 445.

J.C.R. HUNT \& W.E. WILLIAMS 1968 Some Electrically Driven Flows in
Magnetohydrodynamics. Part 1. Theory.\textit{J. Fluid. Mech. }\textbf{31(4) }%
705-722.

J.C.R.HUNT \& S.SHERCLIFF 1971 Magnetohydrodynamics at High Hartmann Number 
\textit{Ann. Rev. Fluid. Mech.} \textbf{3 }37-62.

A.B.TSINOBER \& Y.B.KOLESNIKOV 1974 Experimental Investigation of
Two-Dimensional Turbulence Behind a Grid \textit{Isv. Akad. Nauk. SSSR Mech.
Zhod. i Gaza }\textbf{4 }146.

O.LIELAUSIS 1975 Liquid Metal Magnetohydrodynamics \textit{Atomic Energy
Review }\textbf{13 }527.

R.MOREAU 1990 Magnetohydrodynamics \textit{Kluwer Academic Publishers}

R. MOREAU 1998 \textit{Applied Scientific Research} Magnetohydrodynamics at
the Laboratory Scale : Established Ideas and New Challenges \textbf{58 }%
131-147

P.H.ROBERTS 1967 Introduction to Magnetohydrodynamics \textit{Longmans.}

S.\ SHERCLIFF\ 1953 \textit{Proc. Camb. Phil. Soc.}\textbf{49} 136.

J.SOMMERIA \& R.MOREAU 1982 Why, How and When MHD Turbulence Becomes
Two-Dimensional \textit{J. Fluid. Mech. }\textbf{118 }507-518.

J.SOMMERIA 1988 Electrically Driven Vortices in a Strong Magnetic Field 
\textit{J. Fluid. Mech \textbf{189 }553-569.}

O.ZIGANOV \& A.THESS 1998 Direct Numerical Simulations of Forced MHD
Turbulence at Low Magnetic Reynolds Number \textit{J. Fluid. Mech. }\textbf{%
358 }299-333.

M.I LOFFREDO 1986 Extension of Von Karman Ansatz to Magnetohydrodynamics 
\textit{Mecanica }\textbf{21 }81-86.

H.G.LUGT 1996 Introduction to Vortex Theory \textit{potamac }Maryland.

R.S.NANDA \& H.K. MOHANTY 1970 Hydrodynamic Flow in Rotating Channel \textit{%
Appl. Sci. Res.} \textbf{24 }65-78.

B.M\"{U}CK, C.G\"{U}NTHER, U.M\"{U}LLER \& L.B\"{U}HLER 2000
Three-Dimensional MHD Flows in Rectangular Ducts with Internal Obstacles.%
\textit{submitted to J. Fluid. Mech.}

\end{document}